\documentclass[11pt,letterpaper]{article}
 \pdfoutput=1

\usepackage{jcappub}

\usepackage{caption}
\usepackage{floatrow}
\usepackage{journals}

\usepackage{feynmp}

\usepackage{subfig}

\usepackage{bm}

\def\be{\begin{eqnarray}}
\def\ee{\end{eqnarray}}

\def\k{\bm{k}}

\def\x{\bm{x}}
\def\q{\bm{q}}
\def\s{\bm{s}}

\def\v{\bm{v}}

\newsavebox{\tempbox}

\title{Solving Large Scale Structure in Ten Easy Steps with COLA}
\author{Svetlin Tassev$^{a,b}$, Matias Zaldarriaga$^{c}$, Daniel Eisenstein$^{b}$}
\affiliation{ \sl $^{a}$ Department of Astrophysical Sciences, Princeton University, 4 Ivy Lane, Princeton, \\NJ 08544, USA\\
\sl $^{b}$ Center for Astrophysics, Harvard University, 60 Garden Street, Cambridge, \\MA 02138, USA\\
 \sl $^{c}$ School of Natural Sciences, Institute for Advanced Study, Olden Lane, Princeton, \\NJ 08540, USA
}

\abstract{We present the COmoving Lagrangian Acceleration (COLA) method: an N-body method for solving for Large Scale Structure (LSS) in a frame that is comoving with observers following trajectories calculated in Lagrangian Perturbation Theory (LPT). 
Unlike standard N-body methods, the COLA method can straightforwardly trade accuracy at small-scales in order to gain computational speed without sacrificing accuracy at large scales. This  is especially useful for cheaply generating large ensembles of accurate mock halo catalogs required to study galaxy clustering and weak lensing, as those catalogs are essential for performing detailed error analysis for ongoing and future surveys of LSS.
As an illustration, we ran a COLA-based N-body code on a box of size 100$\,$Mpc$/h$ with particles of mass $\approx 5\times 10^9M_\odot/h$.  Running the code with only 10 timesteps was sufficient to obtain an accurate description of halo statistics down to halo masses of at least $10^{11}M_\odot/h$. This is only at a modest speed penalty when compared to mocks obtained with LPT. A standard detailed N-body run is orders of magnitude slower than our COLA-based code. The speed-up we obtain with COLA is due to the fact that we calculate the large-scale dynamics exactly using LPT, while letting the N-body code  solve for the small scales, without requiring it to capture exactly the internal dynamics of halos. Achieving a similar level of accuracy in halo statistics without the COLA method requires at least 3 times more timesteps than when COLA is employed. 
}

\notoc
\usepackage{graphicx}
\begin{document}
\maketitle

%\vspace{2mm}
\section{Introduction}

Our current cosmological model has introduced two major outstanding puzzles, those of dark matter and dark energy. A lot of effort is put into observational programs to investigate the properties of that dark sector. However, 
extracting unbiased information from future surveys of Large Scale Structure (LSS), such as BigBOSS, WFIRST, Euclid and LSST, relies on our ability to robustly model  structure formation not only in the perturbative linear regime (maximum wavenumber $k_{\mathrm{max}}\sim 0.05h/$Mpc at redshift $z=0$) but also in the non-linear (NL) regime ($k_{\mathrm{max}}\gtrsim k_{\mathrm{NL}} \sim 0.3h/$Mpc at $z=0$), as well as in the intermediate mildly non-linear (MNL) regime.

Ideally, the best way to study the dark sector would be to track the evolution of the density field of Cold Dark Matter (CDM) in real space. The reason for that is that any ``contamination'' in the linear and MNL regimes from complicated-to-model non-virialized NL structures is strongly suppressed due to conservation of mass and momentum, which cause any transfer of power from NL scales to larger scales to go as $(k_{\mathrm{large\  scales}}/k_{\mathrm{NL}})^4$ \cite{PeeblesBook}; while virialized structures have a vanishing effect on large scales \cite{2010arXiv1004.2488B}.

Recently significant progress has been made towards modelling the MNL regime of LSS  \cite{2012JCAP...04..013T,2012arXiv1203.5785T} precisely for such ``ideal'' observations of the CDM in real space using modifications to Lagrangian Perturbation Theory (LPT). Having such accurate models  is an extremely powerful asset. If one is able to push our understanding of LSS from the linear to the MNL regime, then the number of Fourier modes (which is $\sim k_{\mathrm{max}}^3\times$(survey volume)) available for inferring cosmological parameters without introducing systematic errors increases $\sim$100-fold. Thus, if one relied only on linear theory, one would have to perform observational surveys with two orders of magnitude larger volume to capture an equivalent number of modes to the ones available if one understood the MNL regime. Therefore, in  \cite{2012JCAP...04..013T,2012arXiv1203.5785T,2012arXiv1203.6066T}  those models were used to show how one can significantly reduce sample variance in statistics extracted from N-body simulations, as well as to quasi-optimally reconstruct the Baryon Acoustic Oscillations (BAO) peak, which is shifted and smeared at the percent level by MNL evolution  (e.g. \cite{2012arXiv1202.0090P}).

However, instead of such ``ideal'' observations of the CDM density in real space, we usually observe tracers (such as galaxies) of that field in redshift space. Redshift distortions, and the Finger-of-God effects in particular,  couple the NL motions inside galaxy clusters with the MNL scales, while galaxy bias modifies the underlying CDM power spectrum in a time-, scale-, and mass-dependent way  on those scales, including a non-deterministic component. Therefore, to make contact with real-world observations, we need to model the NL regime along with the MNL regime.

In principle, one can get a handle on the MNL and NL regimes by using N-body simulations. However, quantifying  uncertainties requires reducing sample variance for the 4-point function (covariance matrices) of the tracer field. That requires running hundreds and even thousands of simulations, which so far has been feasible only after introducing extremely crude approximations (e.g. the PTHalos method \cite{2002MNRAS.329..629S, 2012arXiv1203.6609M}). Those approximations rely on ad hoc procedures for assigning halo masses, positions and velocities that are known to affect the halo bias, power spectrum, and covariance matrices at the tens-of-percent level \cite{2012arXiv1203.6609M}, with possibly even larger systematics due to redshift-space distortions.  Thus, going beyond the perturbation theory prescription employed by \cite{2012arXiv1203.6609M} for generating mock halo and galaxy catalogs is essential for current and future LSS surveys. 

At this point, one may ask why we are not simply applying the method described in \cite{2012JCAP...04..013T} to reduce the sample variance of statistics extracted from a single mock catalog in order to achieve the same results one obtains by running an ensemble of mocks. The main reason for that is that we need to perform a detailed error analysis on the tracer field \textit{after} processing it with a complicated pipeline, such as BAO reconstruction (e.g. \cite{2012arXiv1203.6066T,2012arXiv1202.0090P}). Such a reconstruction pipeline applies a non-linear transformation to the tracer field in order to remove the effects of MNL evolution and reconstruct the linear CDM density field, thus producing a more robust determination of the acoustic scale. 
That pipeline is non-trivial to incorporate in the method of \cite{2012JCAP...04..013T}, especially when realistic survey boundaries in redshift space are taken into account. Those issues, combined with the fact that BAO reconstruction is inherently imperfect, imply that a proper error analysis requires a covariance matrix, extracted from an ensemble of  processed realistic mock catalogs. \textit{If} one already had such an ensemble, \textit{then} one could use the method of \cite{2012JCAP...04..013T} to reduce the sample variance of the statistics of the tracer field extracted from a single much more detailed high-force-resolution simulation. In that case, the ensemble of mocks would play the role that LPT played in  \cite{2012JCAP...04..013T} in building estimators with suppressed sample variance.

So, we are posed with the problem that numerical simulations are too expensive computationally to allow us to fully quantify the errors from current and future LSS surveys, while perturbative techniques do not give us enough level of realism. A question we need to ask then is whether it is possible to get the best of both worlds by augmenting the results from LPT (or its modifications), which describes well the MNL regime of CDM dynamics in real space \cite{2012JCAP...04..013T,2012arXiv1203.5785T}, with non-perturbative results, which can be efficiently obtained through N-body simulations. 

%Something has to give in such a case. All of the problems above can in principle be addressed with optimized N-body codes which capture correctly the halo statistics without solving exactly for the internal halo dynamics. 

As an answer to that question, we were successful in combining perturbation theory with numerical simulations by developing an easy-to-implement modification to N-body codes, which requires only as few as ten timesteps to capture accurately CDM halo statistics.  We called that modification the COLA (COmoving Lagrangian Acceleration) method, and it is the main focus of this paper.

In Section~\ref{sec:code} we present the general idea behind the COLA method. We give quantitative comparisons between a code using COLA and calculations done with LPT and standard N-body codes in Section~\ref{sec:comp}, and we summarize in Section~\ref{sec:summary}. We leave a detailed description of the COLA algorithm to  Appendix~\ref{app:KDK}. In Appendix~\ref{app:nmax} we provide some considerations about modeling $n$-point function.

\section{Augmenting Simulations with Perturbation Theory}\label{sec:code}

%One route to achieve that is to augment the results from perturbation theory (or its extensions) with  non-perturbative results, which can be efficiently obtained using N-body simulations. 

All N-body codes perform numerous timesteps to obtain a reasonable approximation to structure formation both at large and small scales. However, we know (e.g. \cite{2012JCAP...04..013T}) that the large scales ($\gtrsim 100\,$Mpc$/h$ at $z=0$) are well-described using Lagrangian Perturbation Theory or modifications thereof. The time integration for the large scales in N-body codes does nothing more than simply solving for the linear growth factor. Therefore, if we are to make few timesteps with an N-body code, the large-scale power will be miscalculated only because we would be getting a bad estimate of the growth factor (e.g. left panel of Figure~\ref{fig:cc}). But we do know the exact value of the linear growth factor -- it is given in any textbook dealing with large-scale structure (e.g. \cite{PeeblesBook}).

We therefore seek to decouple the large and small scales in N-body codes by evolving the large scales using exact LPT results and the small scales using a full-blown N-body code. We can recast the equations of motion of CDM in a form suitable for this kind of splitting by going in a frame comoving with ``LPT observers''\footnote{Another N-body and LPT hybrid was proposed in \cite{HHpaper}. In the work that led up to this paper we implemented that hybrid to third order, but found that it is more slowly converging than the COLA method described in this paper.}. Thus, the equation of motion, which schematically is (we reintroduce the omitted constants and Hubble expansion in Appendix~\ref{app:KDK}) 
\be\label{sketchEoM}
\partial_t^2\x=-\nabla \Phi,
\ee
can be rewritten as:
\be\label{sketchEoMLPT}
\partial_t^2\x_{\mathrm{res}}=-\nabla \Phi-\partial_t^2\x_{\mathrm{LPT}}\ , \ \ \hbox{with} \ \x_{\mathrm{res}}\equiv \x-\x_{\mathrm{LPT}}\ ,
\ee
where $\Phi$ is the Newtonian potential calculated from the density field corresponding to particles at positions $\x_{\mathrm{LPT}}+\x_{\mathrm{res}}$;  $\x(\q,t)$ is the Eulerian position of CDM particles at time $t$, which started out at (Lagrangian) position $\q$; $\x_{\mathrm{LPT}}$ is the LPT approximation\footnote{Instead of 2LPT, one can use any other model (such as the models in \cite{2012arXiv1203.5785T}) of CDM particle trajectories as a solution around which to solve with the N-body code. However, we find that simple 2LPT not only works exceedingly well, but also can be easily incorporated in any N-body code at a modest memory cost and effectively zero computational cost.} to $\x$; and $\x_{\mathrm{res}}$ is the (residual) displacement of CDM particles as measured in a frame comoving with ``LPT observers'', whose trajectories are given by $\x_{\mathrm{LPT}}$. Thus, one should think of $\partial_t^2\x_{\mathrm{LPT}}$ as a fictitious force acting on the CDM particles, since we are working in a non-inertial frame of reference.

The standard approach to do the time integration in N-body codes is to discretize the $\partial_t^2$ operator in (\ref{sketchEoM}). The basic idea behind the COLA method is to discretize $\partial_t^2$ only on the left-hand side of eq.~(\ref{sketchEoMLPT}) and to use the exact LPT expression for the fictitious force, $\partial_t^2\x_{\mathrm{LPT}}$.

The equation that COLA solves, eq.~(\ref{sketchEoMLPT}), is manifestly exact. Thus, in the limit of infinite number of timesteps, we recover exactly both the large and small scales just as in the standard approach, eq.~(\ref{sketchEoM}). However, unlike the standard discretized eq.~(\ref{sketchEoM}), as long as we use second order LPT (2LPT) for $\x_{\mathrm{LPT}}$, the COLA equation (\ref{sketchEoMLPT}) also guarantees that the large scales are calculated exactly to second order in the overdensity (i.e. second order in the linear growth factor) for any number of timesteps, even in the case of as few as one timestep. Any errors accrued will be at most third order in the overdensity, as 
the residual displacement $\x_{\mathrm{res}}$ that COLA solves for is third order by construction (when 2LPT is employed for calculating $\x_{\mathrm{LPT}}$).

Since $\x_{\mathrm{res}}$ is third order in perturbation theory, at early times the time dependence of $\x_{\mathrm{res}}$ is simply a superposition of decaying modes as initial conditions for N-body codes are usually calculated in 2LPT. We used that fact in Appendix~\ref{app:modKDK} to write down a modified discretized version of the operator $\partial_t^2$, which is used in COLA. The main implication is that in COLA the residual velocity $\partial_t\x_{\mathrm{res}}$ is optimally decaying within each time step at early times (see  Appendix~\ref{app:modKDK} for details).

As we show below, the COLA method requires only few timesteps to recover accurate halo statistics. Thus, there is a choice to be made of how to optimally distribute those timesteps between the initial and final times. We discuss this in Appendix~\ref{app:KDKALC}.

To explore the idea presented above, we developed COLAcode\footnote{COLAcode is a C code loosely based on the FORTRAN PMcode of A. Klypin and J. Holtzman, found on  \url{http://astro.nmsu.edu/~aklypin/PM/pmcode}}, a serial PM code incorporating the COLA integration method described above (and in detail in Appendix~\ref{app:KDK}). To illustrate the performance of COLA, we ran a simulation with a box size of $100\,$Mpc$/h$, with $256^3$ particles with forces calculated on a PM grid\footnote{Let us briefly explain our choice of PM grid size. If the average comoving interparticle distance is $d$, then the mass per particle is $(\bar \rho d^3)$, where $\bar \rho$ is the mean comoving density. For a halo with $N$ particles (e.g. as detected by a friends-of-friends (FOF) algorithm), the halo mass is $(N\bar \rho d^3)$, which roughly equals $M_\delta = (4 \pi/3) R_\delta^3 \bar \rho\, \delta$, where
$\delta$ is the fractional overdensity at which the halos stop percolating. Thus, we find 
\be
R_\delta/d = \left(\frac{4 \pi\delta}{3N}\right)^{-1/3}\approx 0.4\times(N/50)^{1/3}\ ,
\ee
where the last equality holds for $\delta=200$. Therefore, the ratio of the halo radius (and hence the force softening) to the mean interparticle distance, should be about $0.4$ for a simulation designed for halo mock catalogs, since one needs at least $\sim 50$ particles per halo to obtain a reasonable radial profile for instance. In our N-body example, we chose a PM grid which is 3 times finer than the mean particle separation, thus satisfying the above requirement.}  of $768^3$. The initial conditions were calculated at $z_{\mathrm{initial}}=9$ using the 2LPTic code\footnote{We used the serial version of the code developed by R. Scoccimarro. It can be found here: \url{http://cosmo.nyu.edu/roman/2LPT/}}.

\begin{figure}[t!]
\flushleft
%\vspace{1.5cm}%%%%%%%
\hspace{0.01\textwidth}\textbf{2LPT}%%%%%
\hspace{0.27\textwidth}\textbf{COLA}%%%%%
\hspace{0.25\textwidth}\textbf{Gadget}%%%%%
\\%%%%%%%
\hspace{0.01\textwidth}\textbf{$\sim$ 3 timesteps}%%%%%
\hspace{0.17\textwidth}\textbf{10 timesteps}%%%%%
\hspace{0.18\textwidth}\textbf{$\sim$ 2000 timesteps}%%%%%
\\%%%%%%%
\vspace{-1.5cm}%%%%%%%
\centering
    \subfloat{\includegraphics[width=0.330\textwidth]{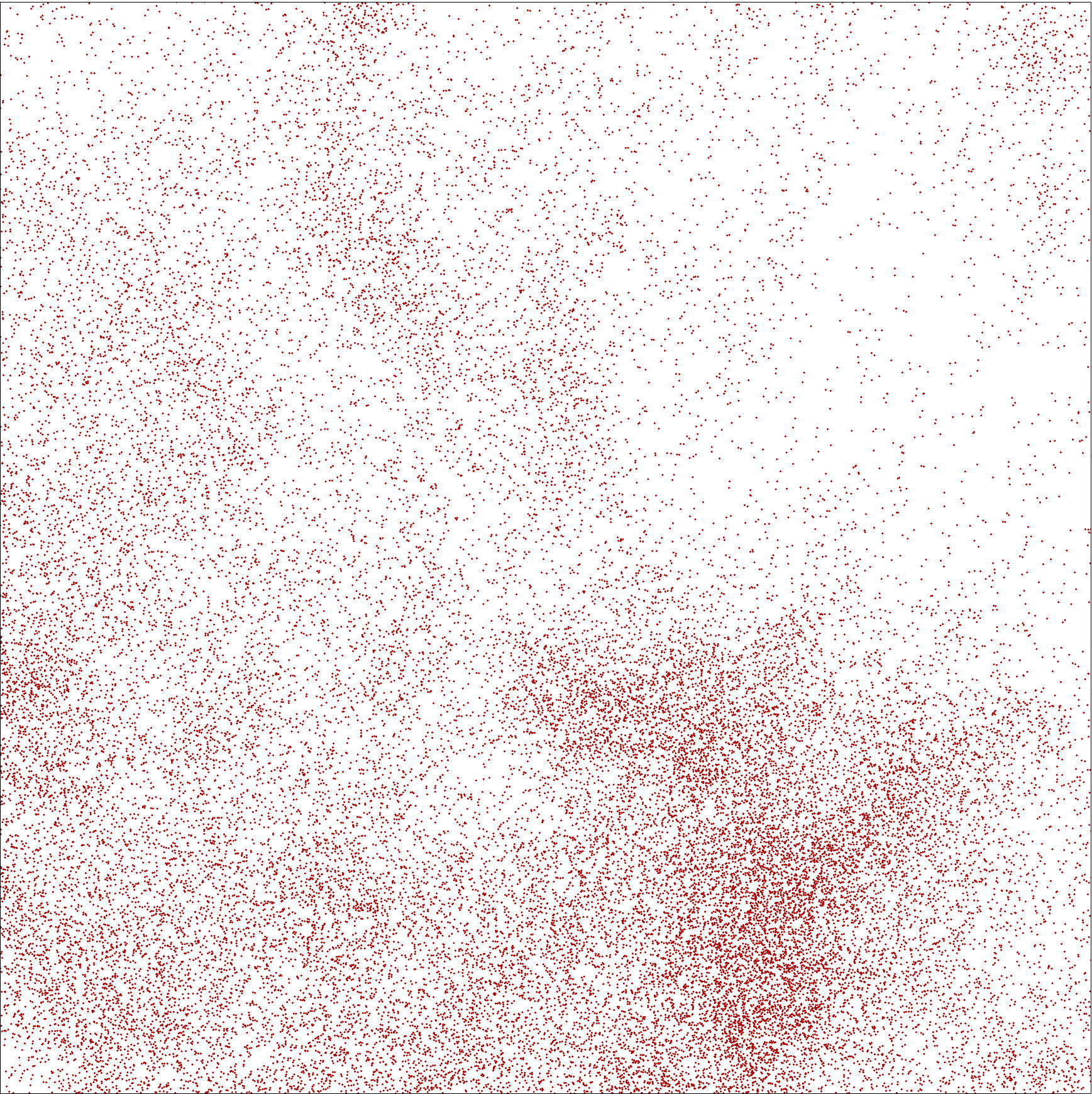}}%%%%%             
  \subfloat{\includegraphics[width=0.330\textwidth]{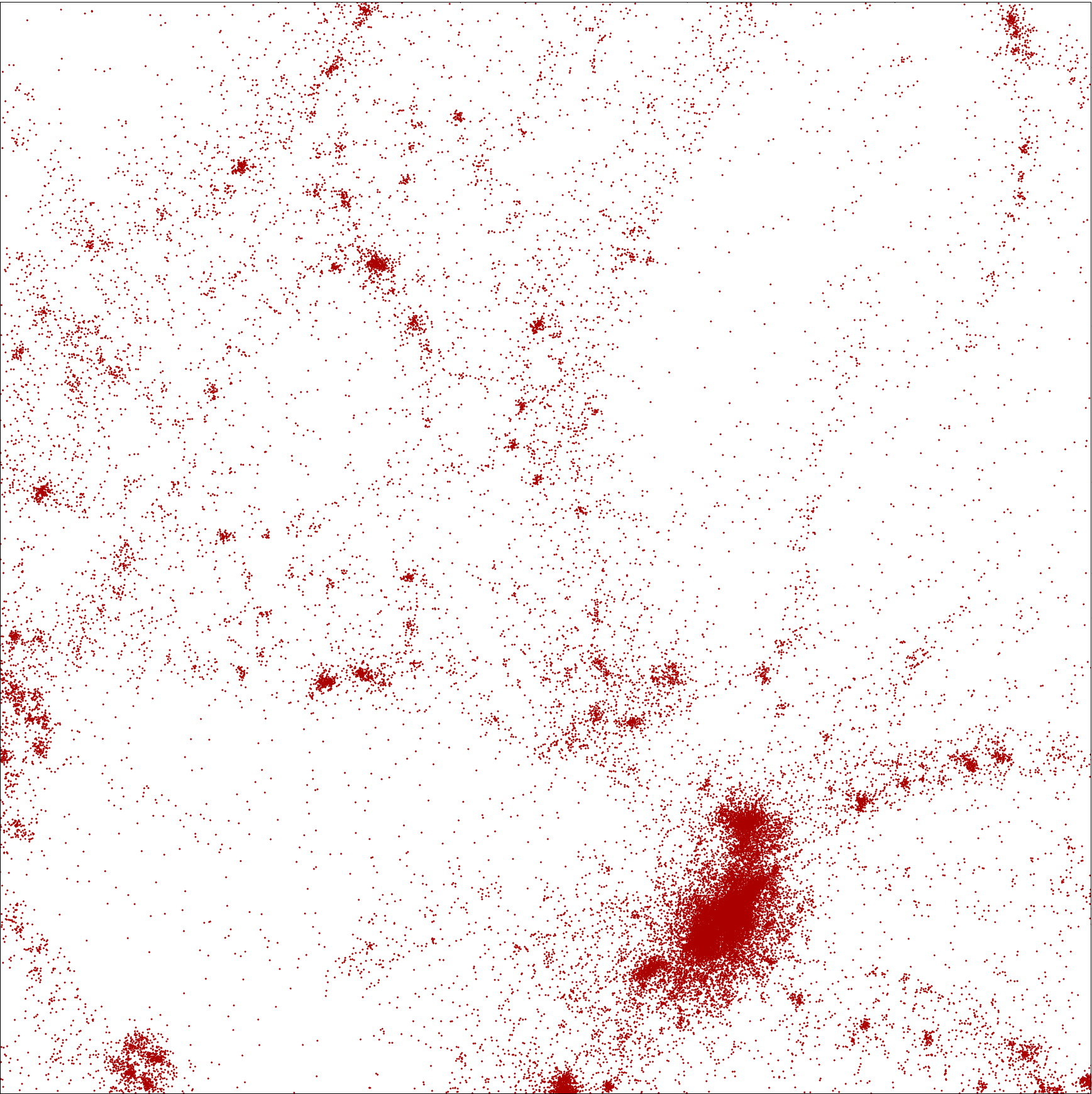}}%%%%%
  \subfloat{\includegraphics[width=0.330\textwidth]{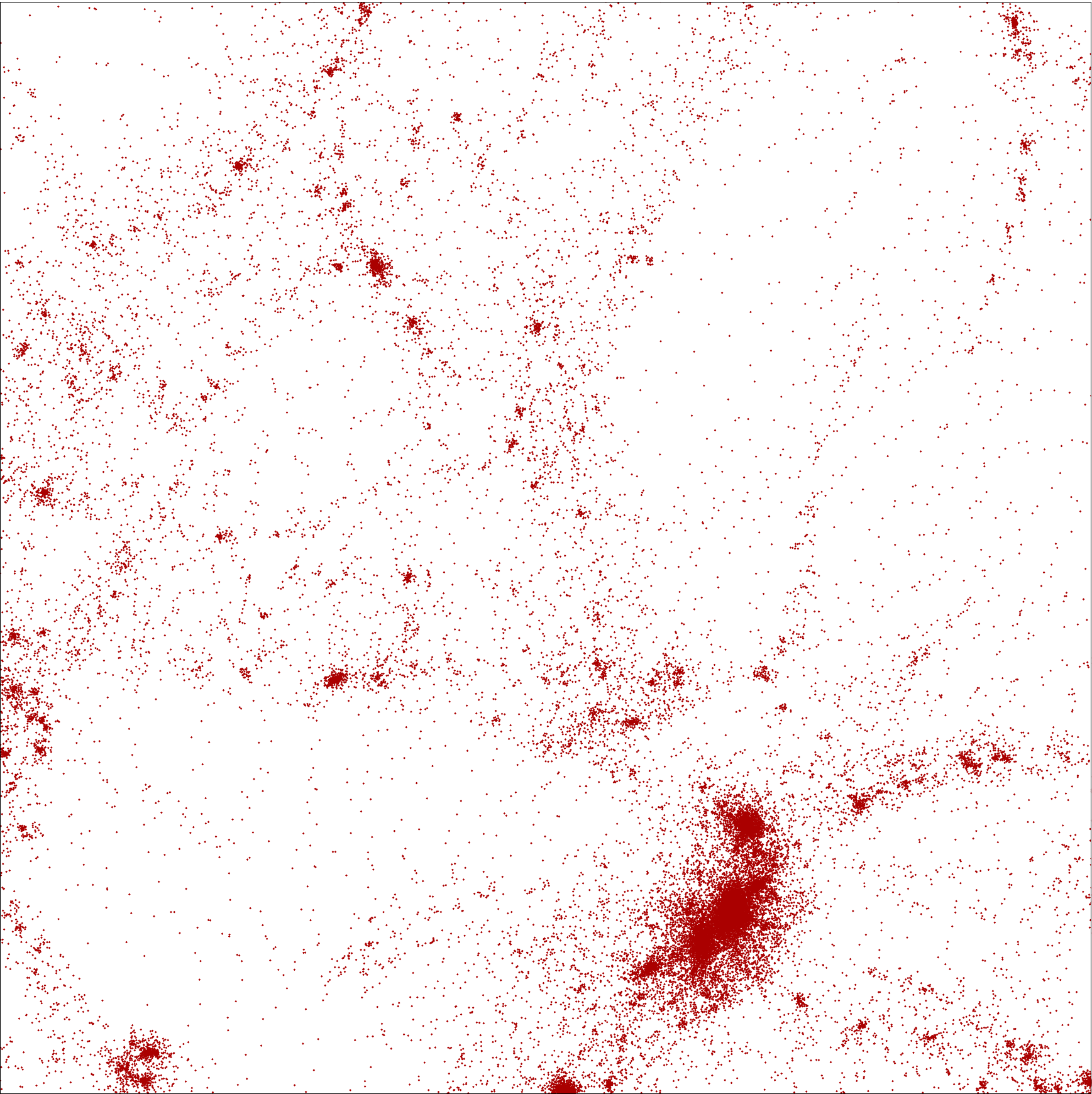}%%%%%
\hfill%%%%%%%
  }%%%%%%%%% 
  \vspace{-1.3mm}
\caption{\small We show slices through three N-body simulations evolving the same initial conditions up to $z=0$. The particles (each of mass $4.6\times 10^9M_\odot/h$) are shown as red points. Each slice is 20\,Mpc$/h$ on the side (the full simulation box is $100\,$Mpc$/h$ on the side), and about 3\,Mpc$/h$ thick. The left panel shows the 2LPT approximation used for building mock catalogs using the PTHalos approach \cite{2012arXiv1203.6609M,2002MNRAS.329..629S}. Calculating the 2LPT particle positions requires an equivalent of roughly 3 timesteps performed by an N-body code. The middle panel shows the result obtained with our modified N-body code with as few as 10 timesteps. The rightmost panel shows the ``true'' result obtained from GADGET-2 \cite{gadget} after $\sim2000$ timesteps starting with 2LPT initial conditions at $z=49$.
\vspace{-4mm}} \label{fig:scatter}
\end{figure}

In Figure~\ref{fig:scatter} we show a slice at $z=0$ through a set of three N-body simulations evolving the same initial conditions. The left panel shows the result performed entirely using 2LPT, which currently used mock catalogs are based upon \cite{2012arXiv1203.6609M}. The central panel shows the snapshot obtained using COLAcode with 10 timesteps; and the right panel shows the result from GADGET-2 \cite{gadget} with $\sim$2000 timesteps.  

The computational cost of our code is only about three times larger than calculating 2LPT initial conditions with standard Fourier techniques \cite{1998MNRAS.299.1097S}, which cost approximately as much as 3 force evaluations in a Particle Mesh (PM) code. The speed-up compared to Gadget that we achieve is entirely due to the fact that we calculate exactly the large-scale behavior in LPT, while letting the N-body code solve for the small-scale dynamics, without requiring it to capture exactly the internal dynamics of halos. 

%Thus, the main advantage of our code is that one can straightforwardly trade accuracy at small-scales in order to gain computational speed, without affecting the accuracy at large scales. To remind the reader, this was the key requirement for generating large ensembles of accurate but cheap mock halo catalogs needed to study LSS.

\section{Quantitative Comparisons}\label{sec:comp}

To quantify the improvement that the COLA method affords us, next we look in detail at how our code (run with 10 timesteps) compares against 2LPT and a PM code run with 10 timesteps based on the standard integration method in Section~\ref{app:KDKstd} (i.e. not running on COLA). In all comparisons we use the optimal COLA variant (see Appendix~\ref{app:KDKALC}) for this simulation setup given by equations (\ref{ALCtimestep},\ref{Lops},\ref{KDKALC}) with timesteps uniform in the scale factor, $a$. 
As a reference (``true'' or ``NL'' quantities) we use a detailed high-force resolution Gadget-2 run starting from the same initial seed but at $z=49$, instead of $z=9$. All comparisons are done in real space at $z=0$.

\subsection{Cross-correlation coefficient and $n$-point functions}

\begin{figure}[t!]
\centering
\hspace{-2mm}%%%
\subfloat{\includegraphics[width=0.365\textwidth]{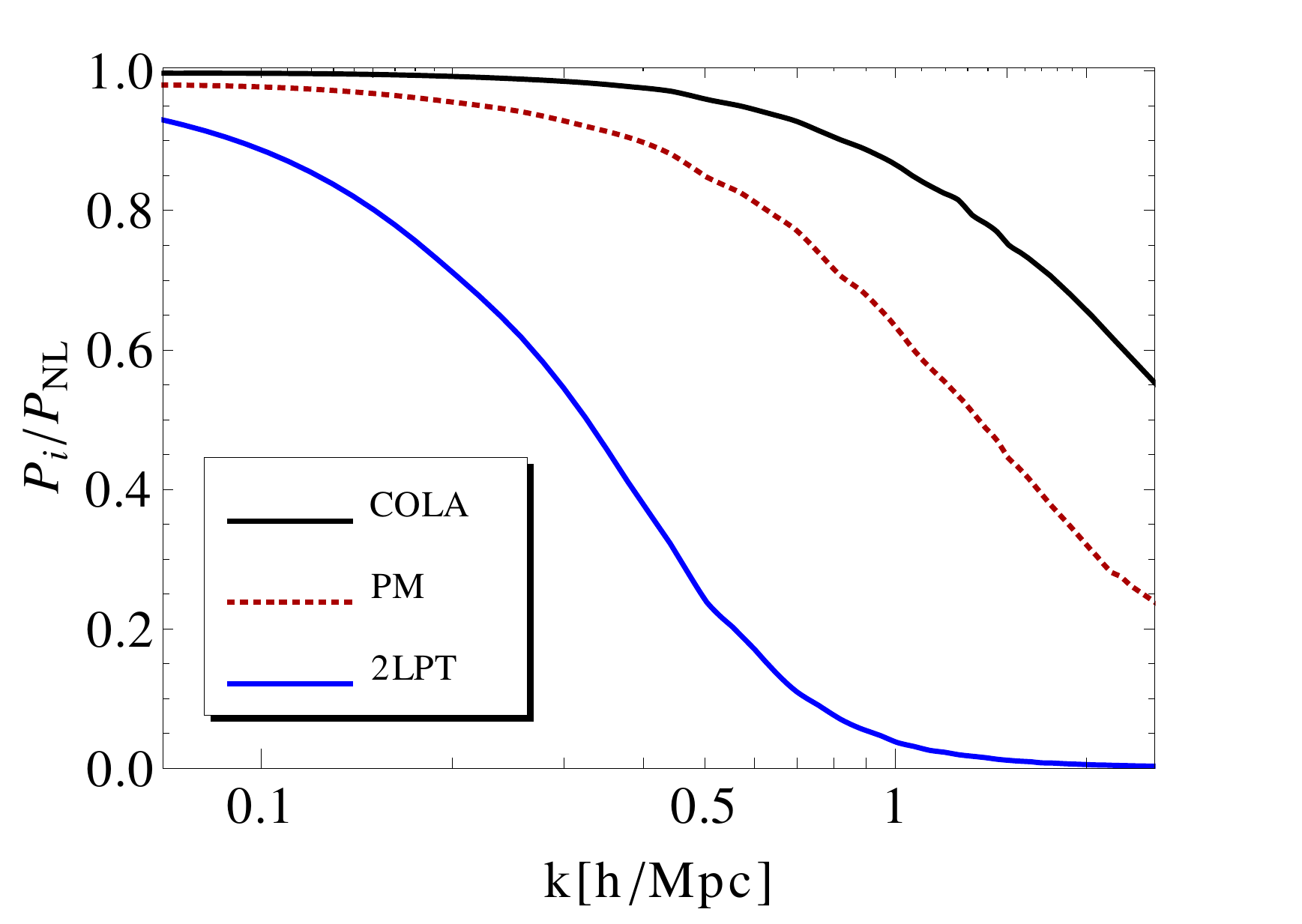}%%%%%
}%%%%
\hspace{-6mm}%%%
\subfloat{\includegraphics[width=0.345\textwidth]{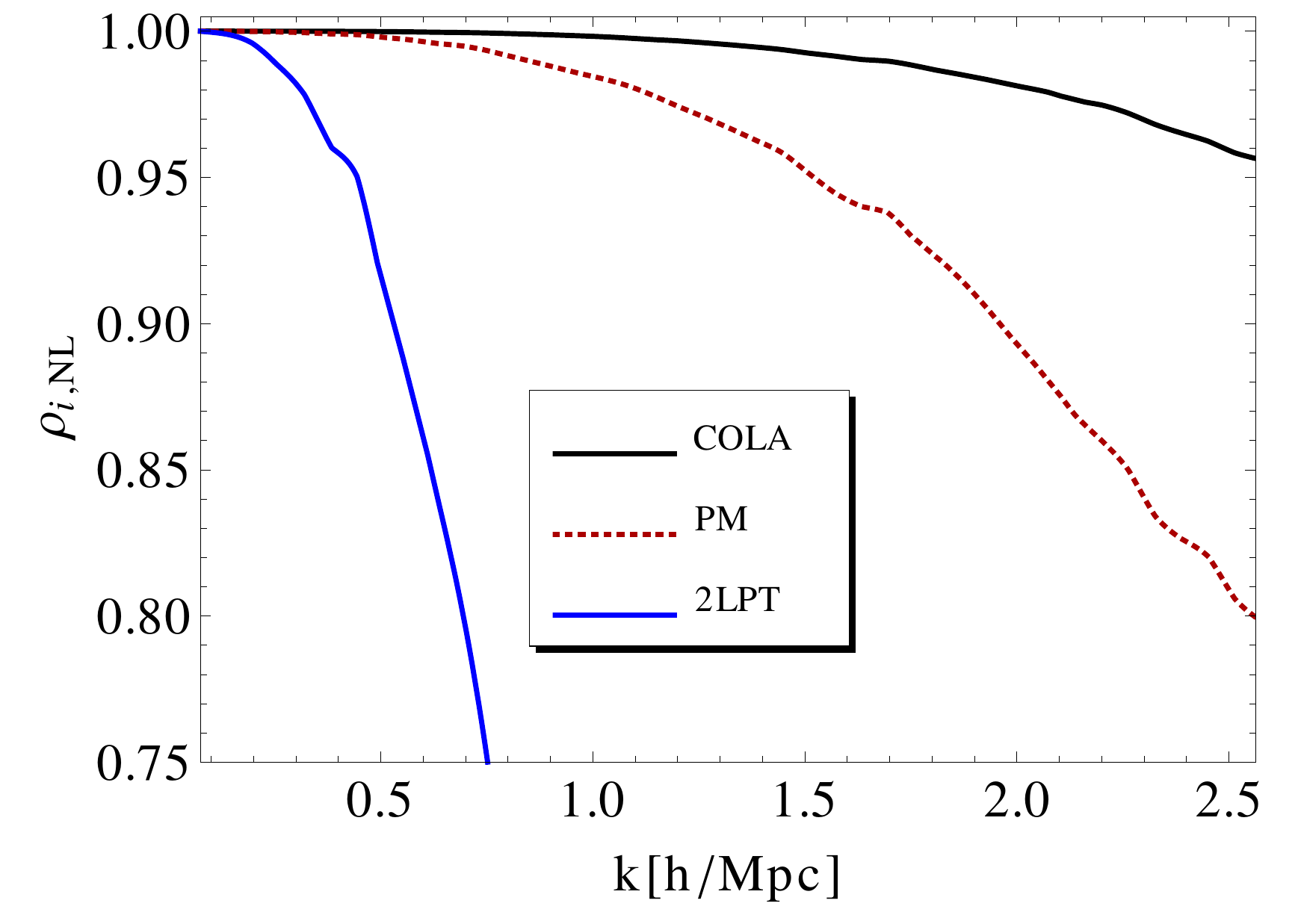}}%%%%%
\hspace{-2mm}%%%
\subfloat{\includegraphics[width=0.35\textwidth]{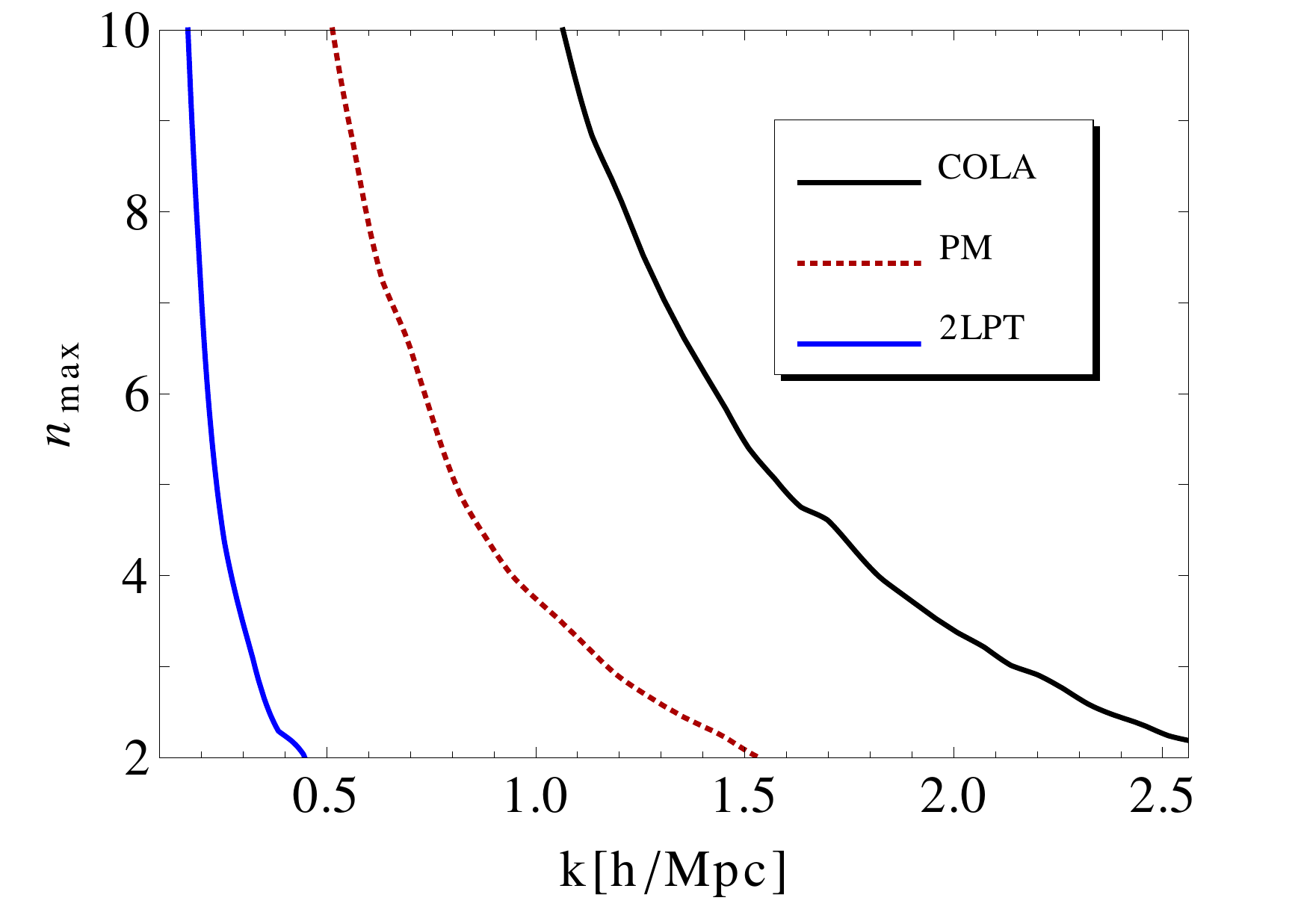}%%%%%%%%%
\hfill%%%%%%%
}%%%%%%%%% 
%  \vspace{-3mm}
\caption{\small We show the power spectrum, cross-correlation coefficient and $n_{\mathrm{max}}$ for a set of models of LSS compared to the ``truth'' (a high-resolution run of Gadget-2). We compare 2LPT currently used for mock catalogs \cite{2002MNRAS.329..629S}, a standard non-COLA PM run with 10 timesteps (``PM''), and the optimal COLA variant again run with 10 timesteps (``COLA''). See the text for further discussion.
%\vspace{-4mm}
} \label{fig:cc}
\end{figure}

In the left panel of Figure~\ref{fig:cc} we plot the ratio of the predicted power spectrum, $P_i$, by the models\footnote{By ``models'' we mean 2LPT or any simulation run that we investigate (e.g. COLA).} ($i$ runs over the models) described above to the ``true'' NL power spectrum. We see that the non-COLA PM code ran with 10 timesteps underestimates the power spectrum at the several-percent level even for the largest scales. As we argued in Section~\ref{sec:code}, this is due to the fact that the few timesteps the code makes are not enough for it to integrate the linear growth factor correctly even for the largest scales. However, as argued in \cite{CC} in principle a simple scale- and time-dependent rescaling can easily solve this problem. Thus, we now focus on robust statistics of the density field.

One of the best ways \cite{CC} to quantify the accuracy of a model of LSS is to use the cross-correlation, $\rho_{i,\mathrm{NL}}$, between the density field, $\delta_i$, the model predicts and the density field, $\delta_{\mathrm{NL}}$, from a detailed high-resolution simulation:
\be
\rho_{i,\mathrm{NL}}(k)\equiv \frac{\langle \delta_i(\k) \delta^*_\mathrm{NL}(\k)\rangle}{\langle |\delta_i(\k)|^2\rangle\langle |\delta_\mathrm{NL}(\k)|^2\rangle}\ ,
\ee
where angular brackets represent ensemble averaging. When we evaluate the above expression numerically, we replace the ensemble averaging by an averaging in $k$-bins. We plot $\rho_{i,\mathrm{NL}}$ in Figure~\ref{fig:cc} for our models: 2LPT; the optimal COLA variant with 10 timesteps; and the non-COLA PM run with 10 timesteps. We can see that COLA achieves excellent cross-correlation with the true result (better than 95\%) out to impressively high $k$ ($>2h/$Mpc) at a modest computational cost compared to 2LPT. COLA gives an improvement of about $80\%$ in scale for a fixed (at a fiducial value of 95\%) cross-correlation coefficient, when compared to a non-COLA PM run with the same number of timesteps. COLA also gives an improvement of more than six times in scale compared to 2LPT at only 3 times the computational cost.

\begin{figure}[t!]
\centering
   \subfloat{\includegraphics[width=0.492\textwidth]{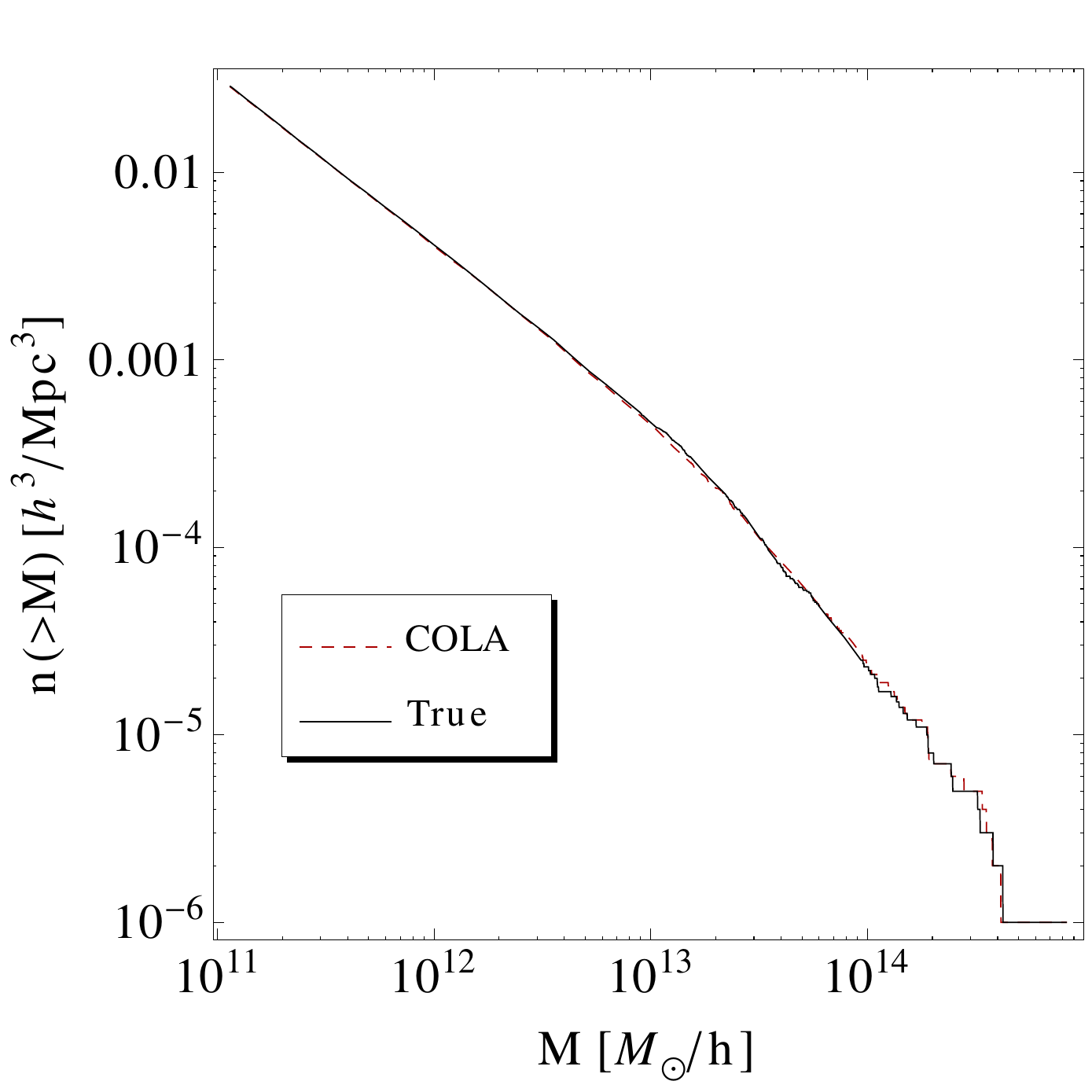}}%%%%%             
\hspace{1mm}%%%
  \subfloat{\includegraphics[width=0.492\textwidth]{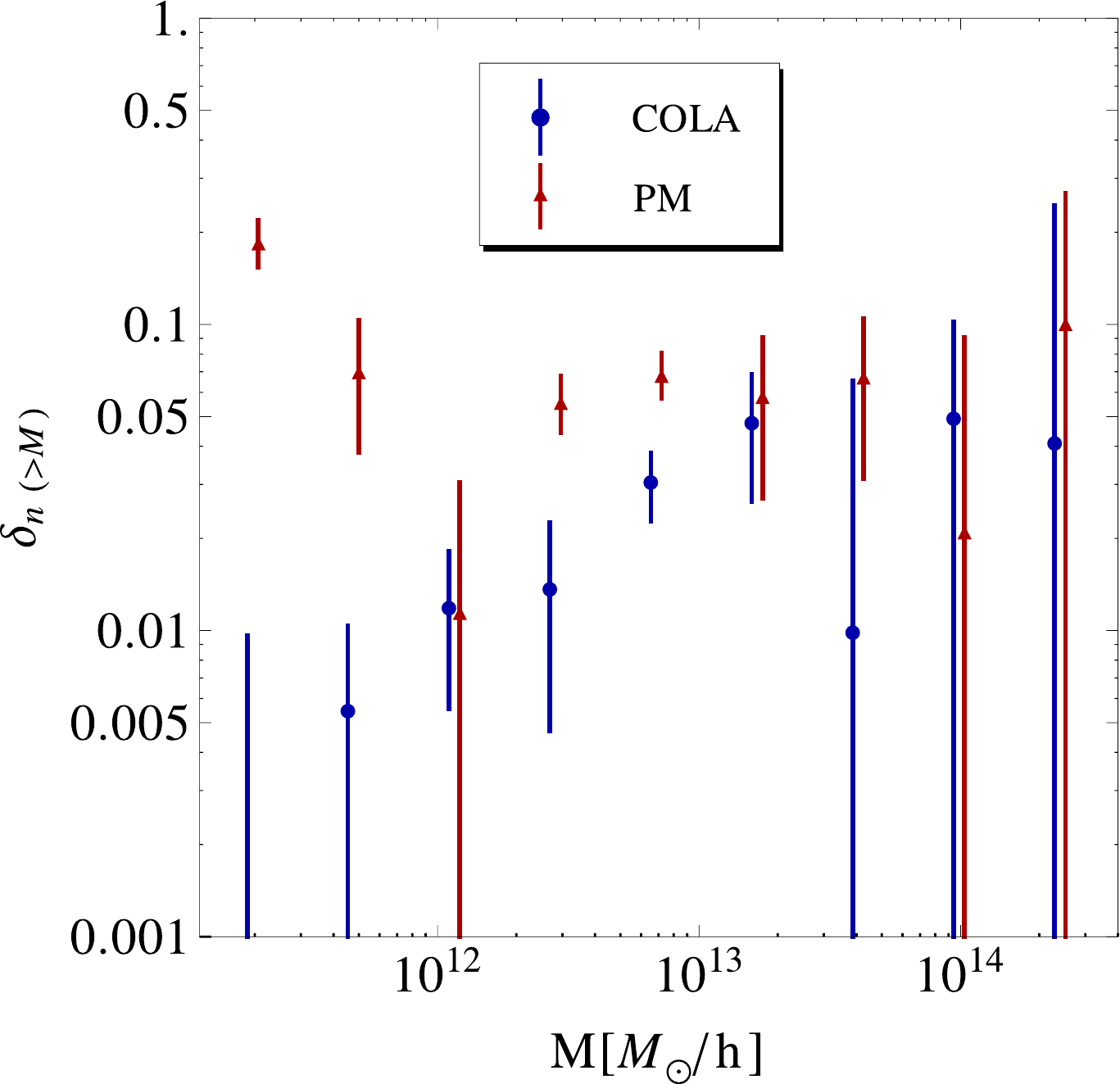}%%%%%
\hspace{1mm}%%%
\hfill%%%%%%%
}%%%%%%%%% 
\caption{\small We show results from running a friends-of-friends halo finder on the $z=0$ output of our improved N-body code (see Figure~\ref{fig:scatter}), when compared against the same results from the Gadget output (denoted by ``True''). The left panel above shows the excellent agreement between the halo mass functions obtained from the results of our COLAcode (with 10 timesteps and a rescaling of the masses by 4\%; see the text for details) and Gadget (with $\sim$2000 timesteps).  The right panel shows the fractional error in the mass function. We include the fractional error for the non-COLA PM run (after a 30\% rescaling of the halo masses to match the high-mass end of the recovered $n(>M)$). Errorbars denote 1-sigma errors in $\delta_{n(>M)}$ obtained in each logarithmic mass bin. Thus, we see that the COLA, non-COLA and Gadget mass functions agree at the 10\% level in the high-mass end. However, the non-COLA PM run still deviates at the 10\% level at the low-mass end, while the COLA run shows about five times better agreement with the true result. 
%\vspace{-4mm}
} \label{fig:Mn}
\end{figure}

The plot of $n_{\mathrm{max}}(k)$ in Figure~\ref{fig:cc} highlights those differences even further. That quantity is defined as
\be\label{nmax}
n_{\mathrm{max}}(k)\equiv \frac{2}{3}\frac{\rho_{i,\mathrm{NL}}}{\sqrt{1-\rho_{i,\mathrm{NL}}^2}}
\ee
and roughly equals the maximum $n$ (as a function of $k$) for which one can trust to $\mathcal{O}(10\%)$ the $n$-point function statistics  for the real-space CDM density field obtained from the particular model at hand. (The derivation of $n_{\mathrm{max}}$ is presented in Appendix~\ref{app:nmax}, but see \cite{CC} for a detailed discussion.) Therefore, we see that for all practical purposes COLA captures correctly the density field down to halo scales. At those scales, the $n$-point statistics are no longer the relevant quantities to look at. We will look at halo statistics next.

\subsection{Halo statistics}

We ran a friends-of-friends (FOF) halo finding algorithm\footnote{We use the University of Washington FOF code available here: \url{http://www-hpcc.astro.washington.edu/tools/fof.html}} on the simulation snapshots presented above. We used a fixed linking length of 0.2 for all snapshots and required at least 25 particles for a halo detection, corresponding to a mass of $1.14\times10^{11}\,M_\odot/h$.

\begin{figure}[t!]
\centering
  \subfloat{\includegraphics[width=0.5\textwidth]{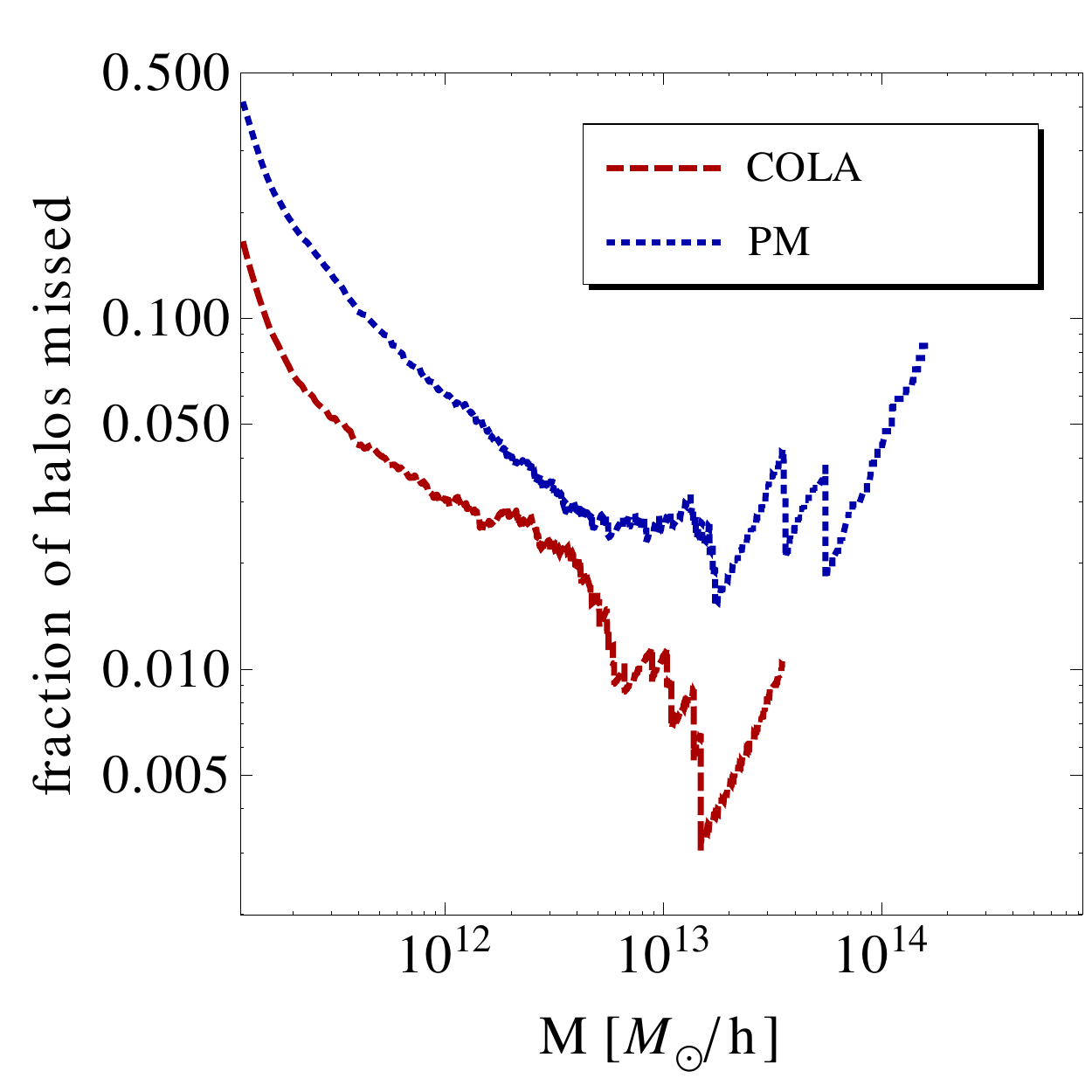}%%%%%
\hfill%%%%%%%
}%%%%%%%%% 
\caption{\small The figure shows the fraction of halos missed by our 1-to-1 matching algorithm, which matches halos between the COLAcode/PM snapshot and the Gadget snapshot.
%\vspace{-4mm}
} \label{fig:missed}
\end{figure}

As we discuss below, we find that halos calculated with COLA with 10 timesteps have slightly smaller masses (as determined by the FOF code) than the true result. This is due to the fact that given the small number of timesteps, gravity is effectively softened, resulting in halos that are generally puffier. We find that this also holds for the non-COLA PM run (with 10 timesteps), although the problem there is much more severe. Below we show that a simple rescaling (by multiplying by a factor of 1.04) of the halo masses works exceedingly well for COLA. The corresponding rescaling for the non-COLA run is much larger (a factor of 1.3 is needed) and only works for a limited range of masses.

The resulting mass function $n(>M)$ is shown in Figure~\ref{fig:Mn}, as well as the magnitude and 1-sigma errorbars for the fractional deviation, $\delta_{n(>M)}$, from the true mass function: $$\delta_{n(>M)}\equiv \left|\frac{n_{i}(>M)}{n_{\mathrm{true}}(>M)}-1\right|\ .$$ As a reference we plot $\delta_{n(>M)}$ as obtained from the non-COLA PM run as well. As mentioned above, the halo masses from the COLA run have been multiplied by 1.04 to match the true halo masses (see also below), while the halo masses from the non-COLA PM run had to be multiplied by 1.3. With that compensating factor,  the non-COLA PM run captures the high-mass end of the mass function correctly but at the expense of the low-mass end, where the error of the mass function is at the  10\% level.

\begin{figure}[t!]
\centering
  \subfloat{\includegraphics[width=0.5\textwidth]{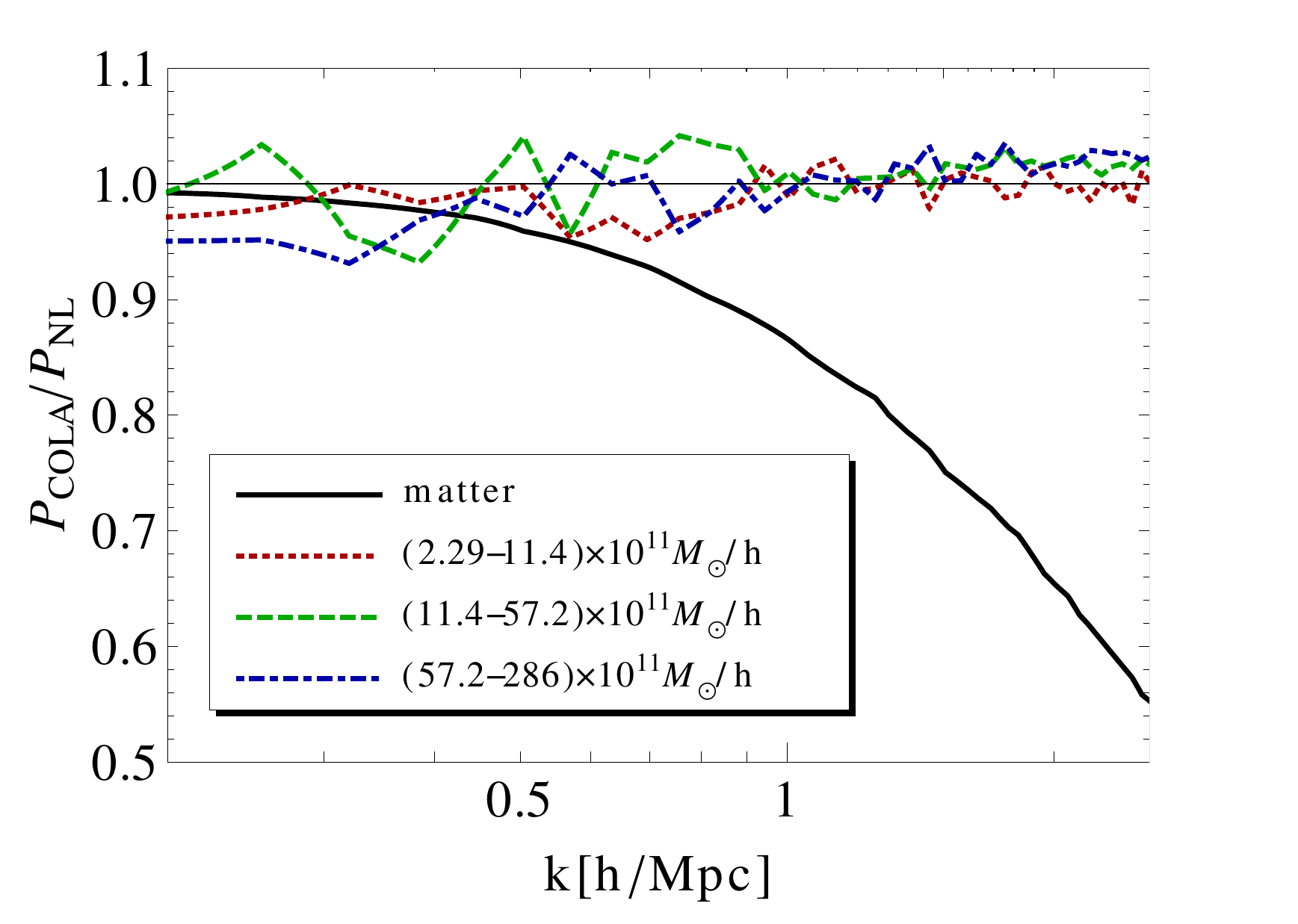}%%%%%
\hfill%%%%%%%
}%%%%%%%%% 
\caption{\small We show the ratio between the COLA and true halo-halo power-spectra in three mass bins (after the correction of 1.04 has been applied to the COLA halos). We include again the ratio between the COLA and true matter-matter power spectra for comparison. We see that COLA recovers the halo power spectrum to within several percent for this particular realization. Since the FOF algorithm detects only halos, and not subhalos, the halo power spectra contain only the 2-halo term, and not the 1-halo term, while the matter power spectrum contains both. Thus, we see that the matter power spectrum deviates from unity because we do not recover the correct 1-halo term. This is due to the fact that COLA halos (after 10 timesteps) are generally puffier than the true ones. See the text for further discussion.
%\vspace{-4mm}
} \label{fig:Phh}
\end{figure}

In Figure~\ref{fig:Phh} we show the ratio between the halo-halo power spectra from the COLA and Gadget snapshots for three halo mass bins. One can see that unlike the corresponding ratio between the matter power spectra, COLA recovers the true halo power spectra to several percent accuracy. To explain that difference, one should note that the FOF algorithm detects only halos and not subhalos. Thus, the halo power spectrum contains only the 2-halo piece and not the 1-halo piece. Since the matter power spectrum contains both, we can conclude that with COLA we do not recover the correct 1-halo term, even though we find the correct distribution of halos. As we will see below, the 1-halo term is not correct since COLA halos (after 10 timesteps) are generally puffier than the truth.

Next we ran an algorithm to detect the 1-to-1 correspondence between halos from the Gadget snapshot on the one hand and the non-COLA and COLA snapshots on the other. The algorithm takes two N-body snapshots which have evolved the same initial conditions. Starting from the heaviest halo in one snapshot the code runs over a given number (in our case 15) of the most bound halo particles trying to find the halo, which most of those particles belong to in the other snapshot. The algorithm is run on both snapshots (thus symmetrized), and the 1-to-1 matched halos are selected. We will compare their properties below. But before we proceed, in Figure~\ref{fig:missed} we show the fraction of halos missed (i.e. halos for which no counterpart is found between the snapshots) in the 1-to-1 matching. Clearly, the COLA method again outperforms the non-COLA PM run by having roughly 4 times fewer missed halos. We can see from that figure that over most of the mass range, the comparisons below cover more than 95\% of the halos.

\begin{figure}[t!]
\centering
  \subfloat{\includegraphics[width=0.324\textwidth]{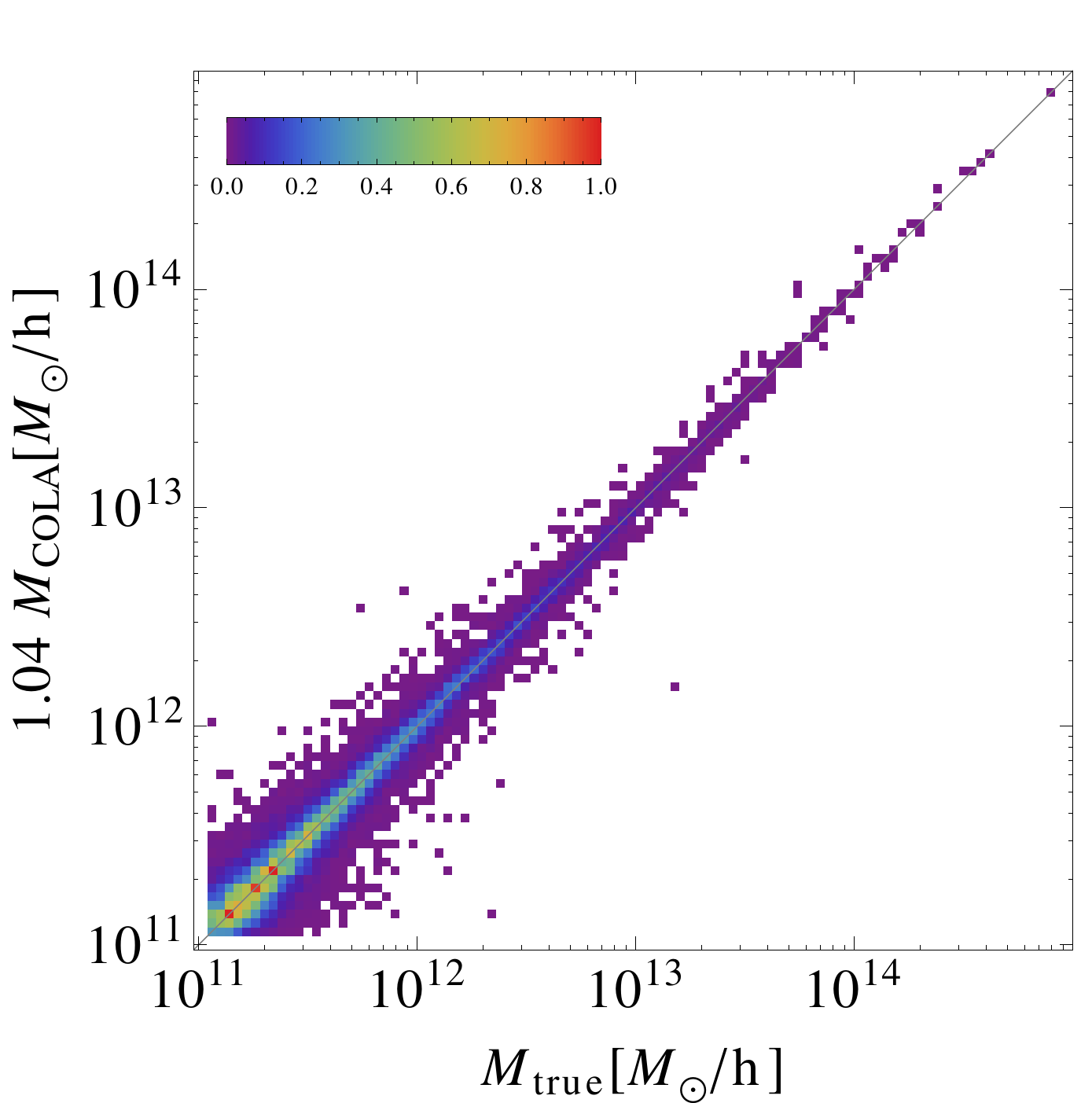}}%%%%%
\hspace{1mm}%%%
  \subfloat{\includegraphics[width=0.324\textwidth]{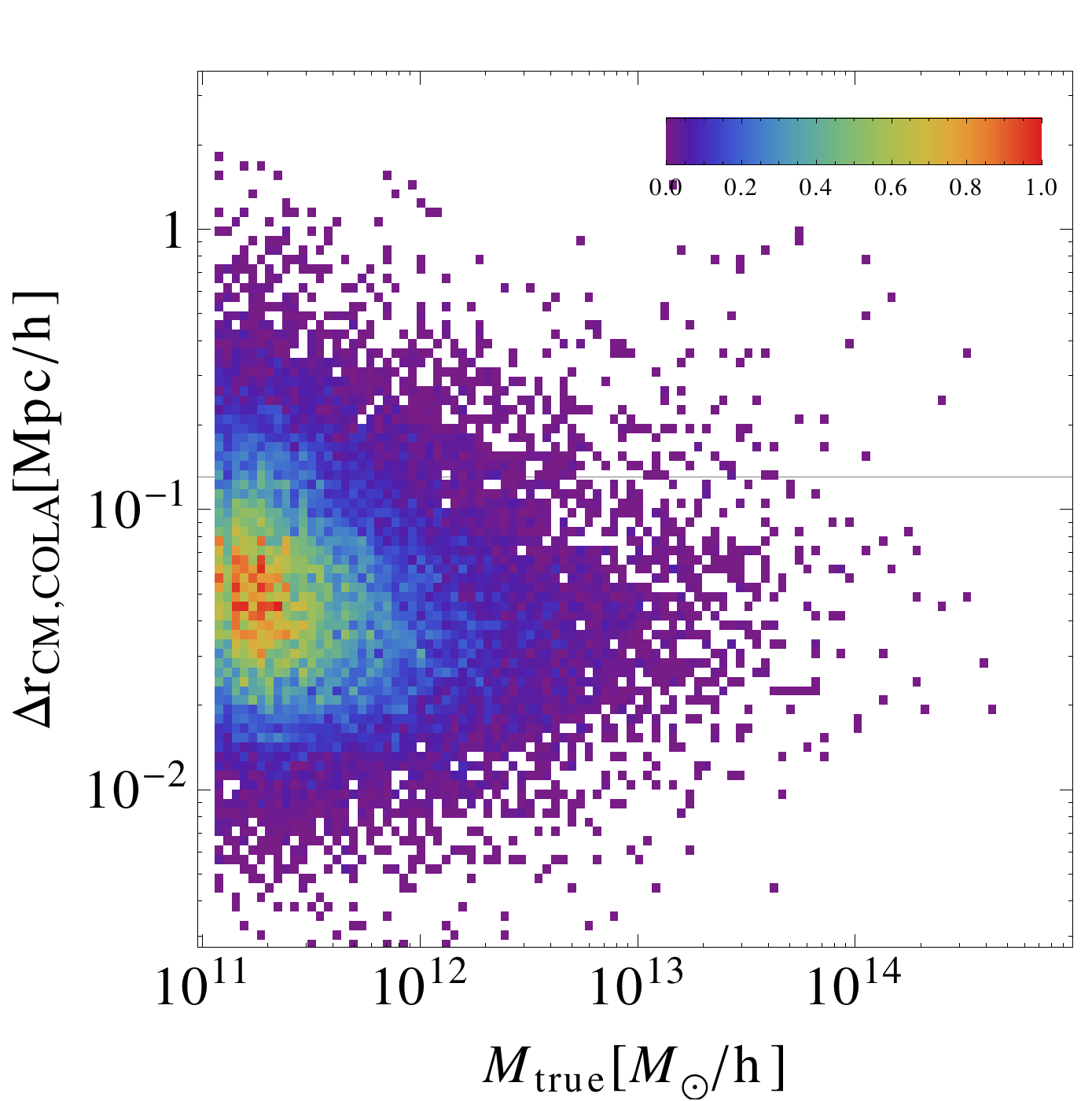}}%%%%%
\hspace{1mm}%%%
  \subfloat{\includegraphics[width=0.324\textwidth]{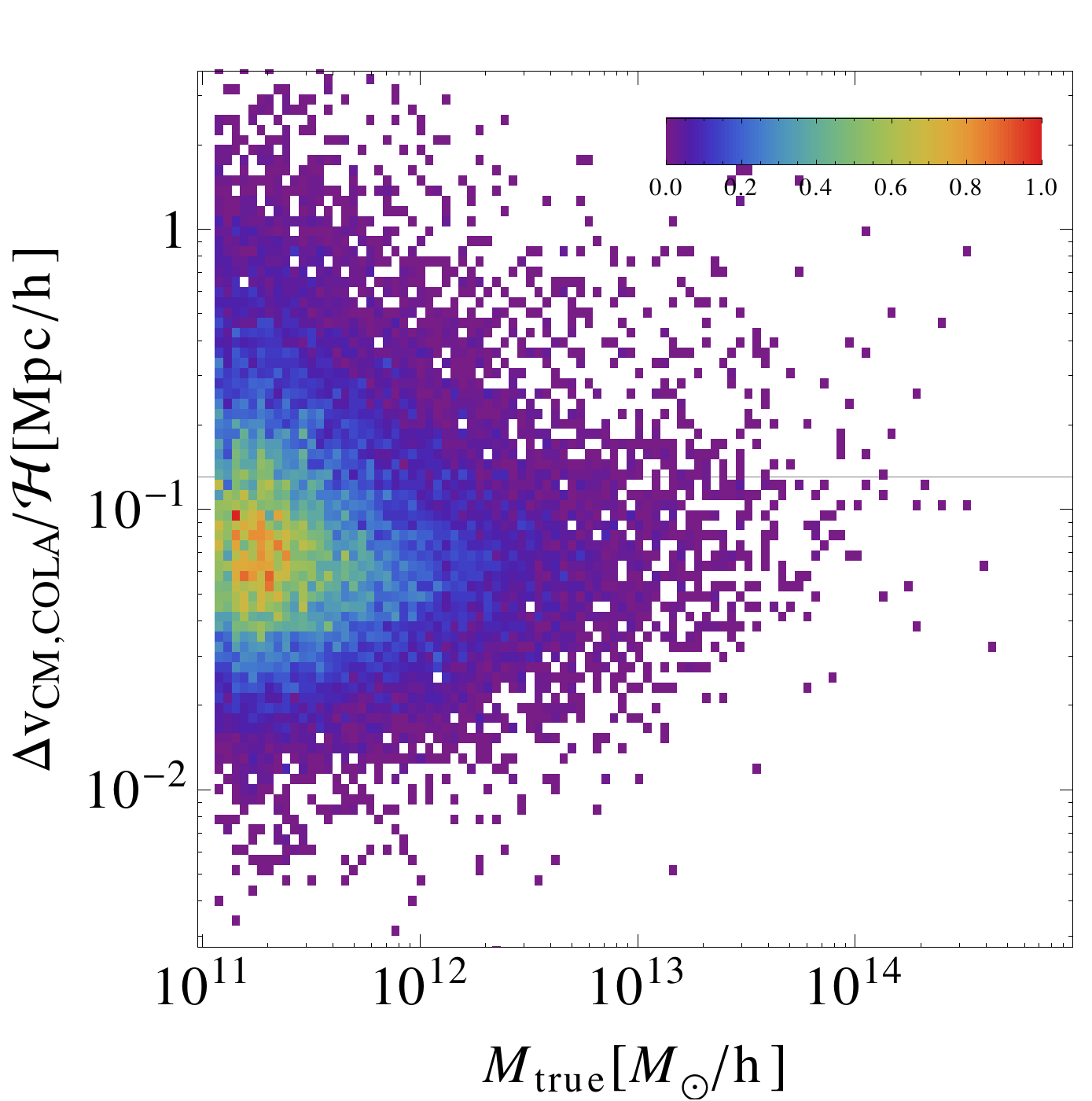}%%%%%
\hfill%%%%%%%
    }%%%%%%%%% 
\caption{\small In the left panel we show the ``True'' mass of each halo against the recovered mass for the same halo from COLAcode. A simple mass rescaling (by 4\%) is enough to bring the correspondence on the gray line indicating the ideal 1-to-1 mass ratio. The middle panel shows the difference distribution in the halo center-of-mass (CM) positions as obtained from Gadget and from our code. The agreement is $\lesssim 100\,$ comoving kpc (gray horizontal line), which corresponds to the force softening length chosen for our code. The right panel shows the difference distribution in the CM velocities in units of Mpc$/h$. Again, those are roughly equal to or better than the force resolution of our simulation. These results indicate that our code is suitable for the fast generation of detailed mock catalogs for studying the clustering of matter on large scales in redshift space. The plot is for $z=0$. The color scale is linear in halo number density per cell in the plot. Here we have used $\mathcal{H}\equiv a H(a)$.
%\vspace{-4mm}
} \label{fig:CM}
\end{figure}

In Figure~\ref{fig:CM} we show the mass correspondence between one and the same halos found in the COLAcode and Gadget snapshots. We also show the difference in center-of-mass positions and velocities for those halos. Due to the larger number of particles, more massive halos have better determined CM positions and velocities. The errors for both those quantities as obtained by COLAcode are roughly equal to or better than the force softening due to the PM grid employed in COLAcode. The fact that the CM velocities are captured so well tells us that the COLA method is well suited for studying halo statistics not only in real space, but in redshift space as well.

\begin{figure}[t!]
\centering
   \subfloat{\includegraphics[width=0.324\textwidth]{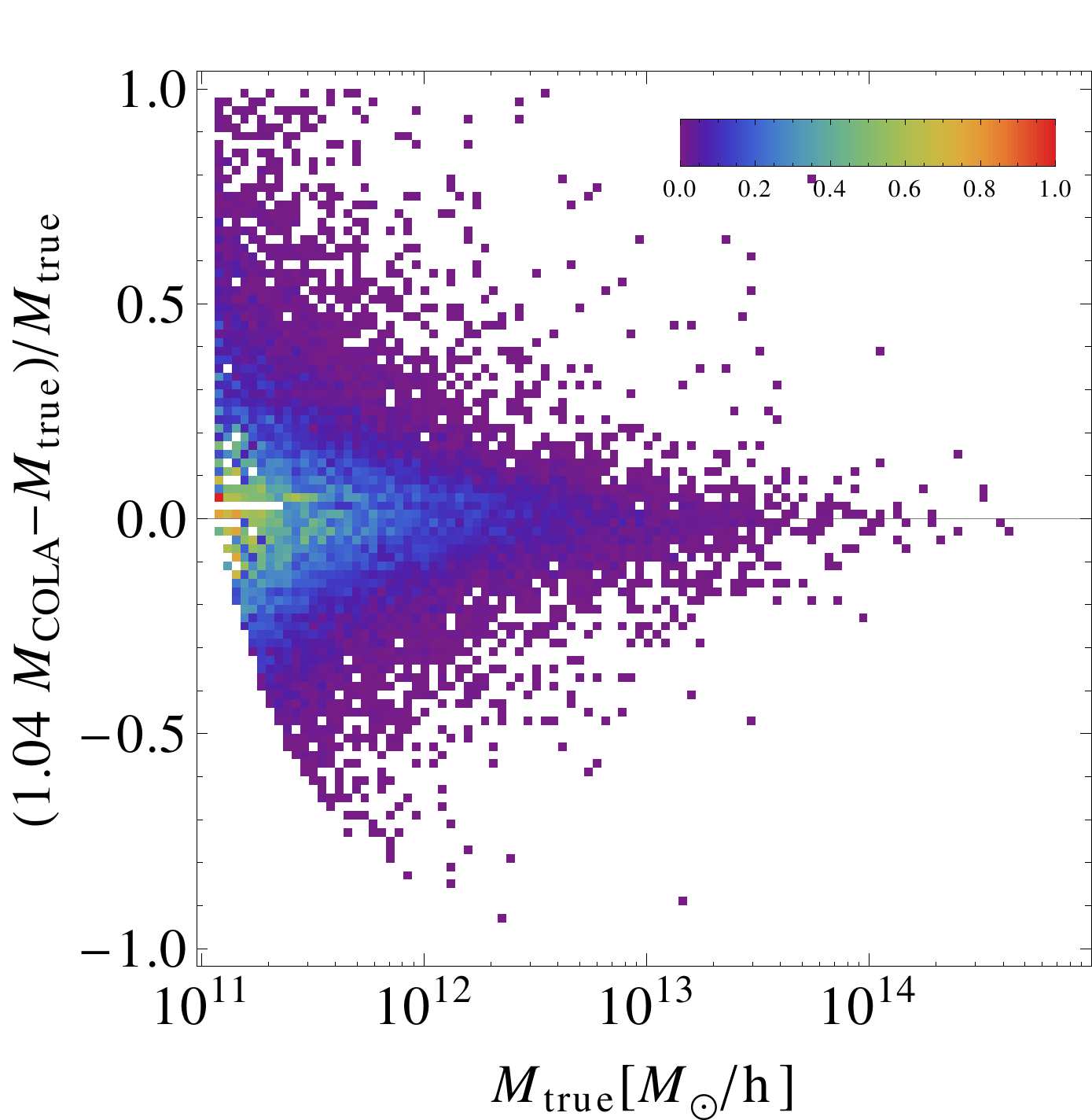}}%%%%%             
\hspace{1mm}%%%
  \subfloat{\includegraphics[width=0.324\textwidth]{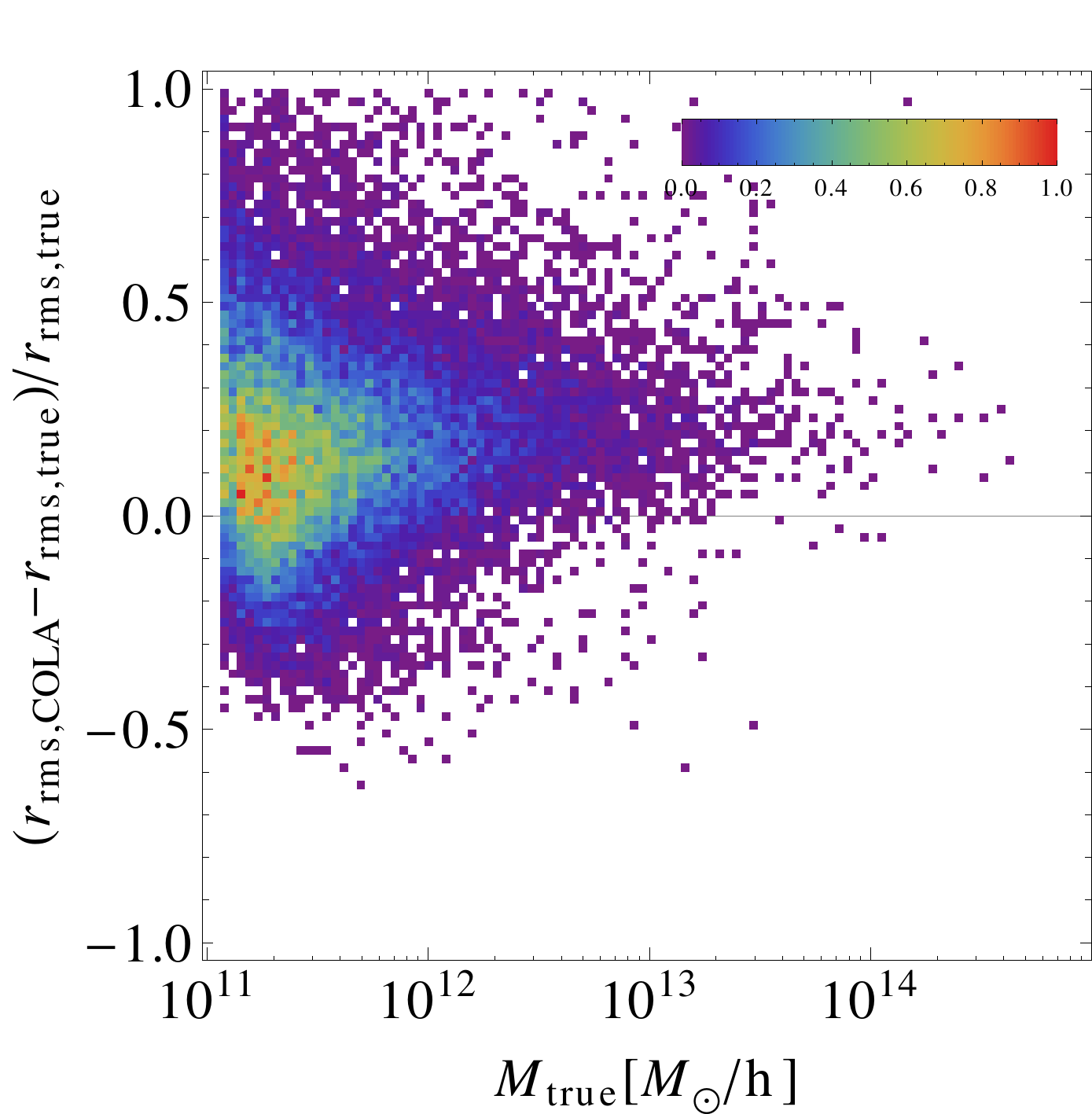}}%%%%%
\hspace{1mm}%%%
  \subfloat{\includegraphics[width=0.324\textwidth]{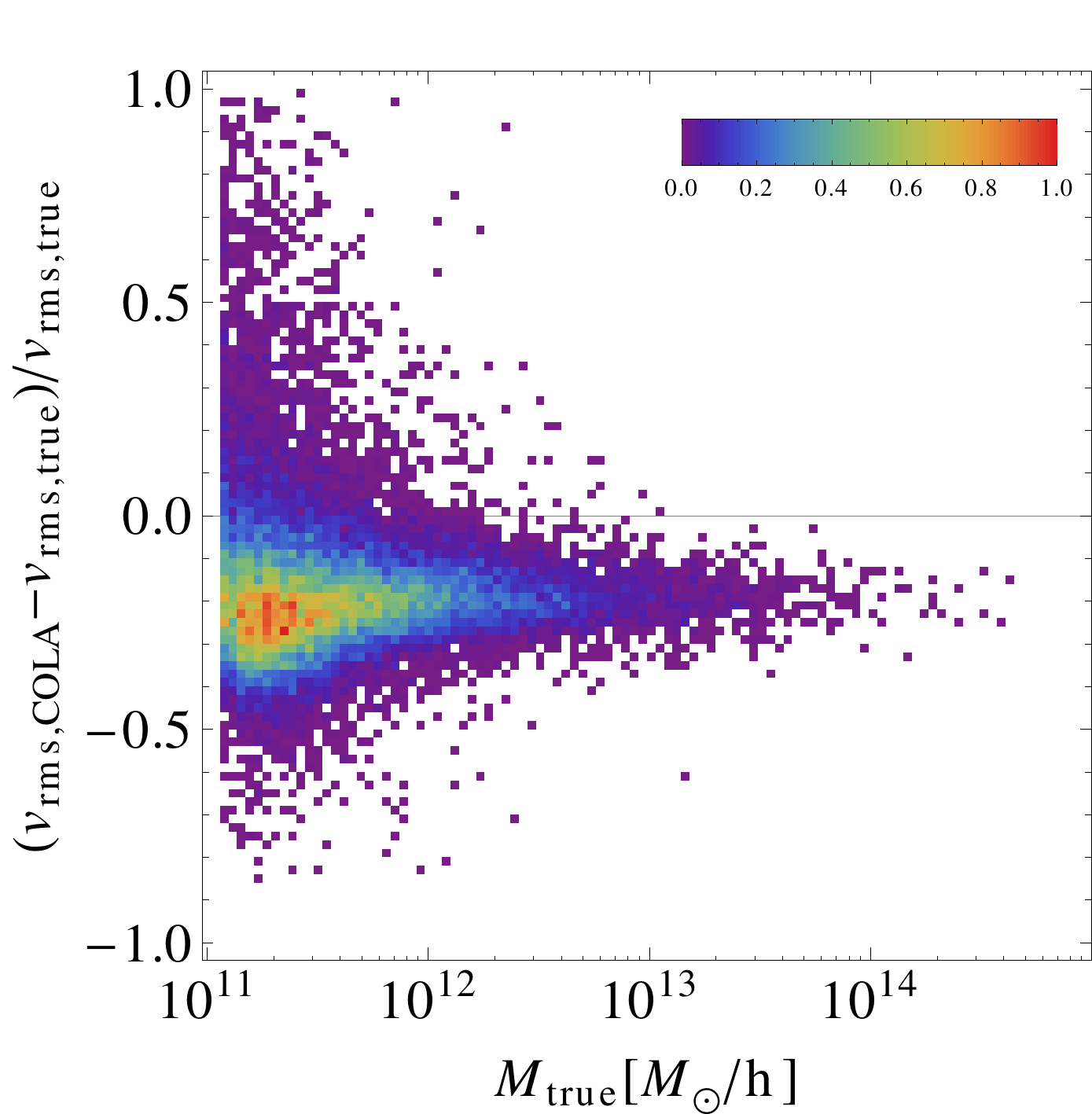}%%%%%
\hfill%%%%%%%
  }%%%%%%%%% 
\caption{\small We show the distribution of fractional errors in the mass, root-mean-square (rms) radius and rms velocity of the halos matched between Gadget and COLAcode. We can see that due to the few timesteps taken by COLAcode, gravity is effectively smoothed, and halos are puffier on average, leading to lower rms velocities. The plot is again for $z=0$. The color scale is linear in halo number density per cell in the plot. The sharp cut in the lower left corner of the left panel is due to the fact that we imposed a minimum of 25 particles per halo in both the COLA and Gadget snapshots.
%\vspace{-4mm}
} \label{fig:ds}
\end{figure}

In Figure~\ref{fig:ds} we show the fractional errors  in the mass, root-mean-square (rms) radius and rms velocity of the halos matched between Gadget and COLAcode. Clearly the 10 timesteps that we take with COLA are not enough to resolve in detail the orbits of particles within halos.  In effect, gravity is softened, and halos are puffier and have lower rms velocities than the real world. This puffiness can explain the 4\% lower halo masses in the COLA snapshots obtained by the FOF finder. Interestingly, not only the mass, but also the rms velocities can be corrected by a  simple multiplicative  rescaling. This opens the door to developing semi-analytical improvements of the recovered halo profiles with COLA, but we leave that for future work.

We already saw that the standard (non-COLA) leapfrog prescription performs worse than the COLA prescription for the same number of timesteps. However, one can ask how many timesteps are needed for the standard prescription to match the quality of the COLAcode 10-timestep runs. The answer is about 30 timesteps for the simulation box studied as an example in this paper.

\section{Summary}\label{sec:summary}

In this paper we presented a straightforward modification (called the COLA method) to N-body codes that forces those codes to solve for the trajectories of CDM particles in a frame comoving with LPT observers. 
With this prescription, the dynamics at large scales are calculated exactly to second order using 2LPT, while the small scales are calculated by the N-body code.

We incorporated the idea presented in this paper in a  PM code, called COLAcode, and explored its advantages over standard (non-COLA) N-body codes. We focused on a comparison with a high-force resolution simulation done with Gadget, with a crude non-COLA run, as well as a comparison with 2LPT results, which have so far been used to construct halo mock catalogs based on the PTHalos method. 

For the illustrative example which we use in this paper, we find that COLAcode is orders of magnitude faster than a high-force accuracy N-body run. This speed-up is dominated by the fact that we are not interested in the detailed particle dynamics inside halos. 

Thus, we also compared COLAcode with a non-COLA PM code. We found that the COLA method is three times faster than a crude non-COLA PM run for the same accuracy for halo statistics (down to halo masses of $10^{11}\,M_\odot/h$). Compared to 2LPT, the COLA method is about three times more computationally expensive. Thus, speed-wise the COLA method  sits  between (on a log-scale) standard N-body codes and perturbation theory. 

Given the few timesteps that COLAcode performs, gravity is effectively softer, and we find that halos are generally puffier and have lower rms velocities. However, we find that halo masses are easily recovered after a simple rescaling (which is $4\%$ for the example we show). We also find that halo center-of-mass positions and velocities are accurate to about $100\,$kpc/h (the force softening of our simulations), which is more than enough for constructing accurate mock halo catalogs in redshift space.

The equation of motion behind the COLA method is exact, as is the case with standard N-body codes. However, unlike those codes, with COLA one can straightforwardly trade accuracy at small-scales in order to gain computational speed, without sacrificing the accuracy at large scales. This crucial advantage is especially useful for cheaply generating large ensembles of accurate mock catalogs required to study galaxy clustering and weak lensing, as those catalogs are essential for performing detailed error analysis for ongoing and future surveys of large scale structure.

Once a realistic ensemble of mock catalogs has been generated with COLA, one can apply the method of \cite{2012JCAP...04..013T} to reduce the sample variance of tracer statistics extracted from a single much more detailed high-force-resolution simulation.  This has numerous possible applications, such as efficiently exploring the cosmological parameter space, training COLA simulations to include unresolved physics, and so on.

COLAcode can be found on the following URL: \url{https://bitbucket.org/tassev/colacode/}

\acknowledgments ST would like to thank Martin White and David Spergel for useful discussions.

\appendix
\section{The details behind COLA}\label{app:KDK}
\subsection{Preliminaries}
One can write the equation of motion for the displacement of CDM particles $\s$ as
\be\label{eom}
T^2[\bm{s}(\bm q,a)]=-\frac{3}{2}\Omega_M a\partial_{\x}\partial_{\x}^{-2}\delta(\x,a)\ , \ \ \mathrm{where}\\
\x(\q,a)=\q+\s(\q,a)\ , \ T\equiv \frac{a}{H_0}\partial_{\eta}= Q(a)\partial_a\ , \ Q(a)\equiv a^3 \frac{H(a)}{H_0}\ .\nonumber
\ee
Here $a$ is the cosmological scale factor; $\eta$ is conformal time; $\Omega_M$ is the usual matter density parameter today; $H$ is the Hubble parameter, with $H_0$ its value today; $\x(\q,a)$ is the comoving (Eulerian) position of the CDM particle, which started out at an initial (Lagrangian) position $\q$; and $\delta$ is the fractional matter overdensity. With these definitions, the (scaled) canonical velocity is given by
\be
\v=T[\s]=\frac{a}{H_0}\frac{\partial \s}{\partial \eta}\ .
\ee

\subsection{The standard leapfrog}\label{app:KDKstd}

The standard leapfrog \cite{1997astro.ph.10043Q} Kick-Drift-Kick algorithm relies on discretizing the time operator $T$ in (\ref{eom}) using the following set of Kick ($\mathrm{K}$) and Drift ($\mathrm{D}$) operators:
\be\label{KDKStd}
\mathrm{D}(a_i,a_f;a_c):& \ \  \x(a_i) \ \mapsto 
\  \x(a_f)=&\x(a_i)+\v(a_c) \int\limits_{a_i}^{a_f} \frac{d\tilde a}{Q(\tilde a)}\nonumber
\\
\mathrm{K}(a_i,a_f;a_c):& \ \ \v(a_i) \ \mapsto  
\ 
\v(a_f)=&\v(a_i)-\frac{3}{2}\Omega_M \partial_{\x}\partial_{\x}^2 \delta(\x,a_c)\int\limits_{a_i}^{a_f} \frac{\tilde a}{Q(\tilde a)}d\tilde a\ ,
\ee
where $a_i$ is some initial and $a_f$ is some final scale factor; $a_c$ is the scale factor for which we have calculated the canonically conjugate variable. 

With the above definitions, the time evolution between $a_0$ and $a_{n+1}$ is then achieved by applying the following operator on $(\x(a_0) , \v(a_0))$:
\be
\prod_{i=0}^n\mathrm{K}(a_{i+1/2},a_{i+1};a_{i+1})\mathrm{D}(a_i,a_{i+1};a_{i+1/2})\mathrm{K}(a_i,a_{i+1/2};a_i)\ .
\ee
When $\Delta a=a_{i+1}-a_i$ is infinitesimal, the optimal choice for $a_{i+1/2}$ is $(a_i+a_{i+1})/2$ as that also corresponds to the mid-point in time for the interval $(a_i,a_{i+1})$. That in turn leads to errors in the particle positions and velocities which are only third order in $\Delta a$.

One has a choice of how to distribute the $a_i$'s. Looking at the cross-correlation coefficient between the true density field and the density field obtained by a PM code using the above KDK scheme, we find that a uniform distribution in $\log (a)$ gives much better results than a uniform distribution in $a$. That is not surprising since it implies a constant timestep in units of the Hubble time.

\subsection{The COLA method}\label{app:KDKALC}

The COLA method dictates that we integrate (\ref{eom}) in a frame comoving with observers following trajectories  specified by LPT. Thus, we solve for the residual displacement\footnote{In \cite{2012arXiv1203.5785T} the displacement residual is called the mode-coupling displacement, $\s_{\mathrm{MC}}$ (as long as we set the transfer functions there to 1). Thus, using that language, what the COLA method does is solve for the non-trivial mode-coupling term numerically.}
\be
\bm{s}_\mathrm{res}\equiv \bm{s}-D_1\bm{s}_1-D_2\bm{s}_2\ ,
\ee
where $D_1(a)$ and $D_2(a)$ are the linear (or Zel'dovich) and second-order growth factors, respectively (normalized such that $D_1(1)=D_2(1)=1$); while $\s_1(\q)$ and $\s_2(\q)$ are the Zel'dovich (linear) and 2LPT displacement fields at $a=1$, respectively.
Thus, the CDM equation of motion can be written as
\be\label{eomALC}
T^2[\bm{s}_\mathrm{res}]=-\frac{3}{2}\Omega_M a\partial_{\x}\partial_{\x}^{-2}\delta(\x,a)-T^2[D_1]\s_1-T^2[D_2]\s_2\ .
\ee
As described in Section~\ref{sec:code}, we discretize the operator $T$ only on the left-hand side of the above equation, using as a velocity variable $\v_{\mathrm{res}}\equiv T[\s_{\mathrm{res}}]$, which is the CDM velocity as measured by an LPT observer. 

As we described in Section~\ref{sec:code}, the COLA method requires only few timesteps to recover accurate halo statistics. Thus, we need to decide on how to optimally 1) discretize $T$, and 2) distribute the timesteps between the initial and final times. 

\subsubsection{Discretizing $T$ in COLA in the standard approach}

We can discretize $T$ following the standard prescription (Section~\ref{app:KDKstd}):
\be\label{KDKALCstd}
\mathrm{D}(a_i,a_f;a_c):& \ \  \x(a_i) \ \mapsto 
\  \x(a_f)=&\x(a_i)+\v(a_c) \int\limits_{a_i}^{a_f} \frac{d\tilde a}{Q(\tilde a)}+\nonumber
\\
&&+(D_1(a_f)-D_1(a_i))\s_1+(D_2(a_f)-D_2(a_i))\s_2\ ,\nonumber
\\
\mathrm{K}(a_i,a_f;a_c):& \ \ \v(a_i) \ \mapsto  
\ 
\v(a_f)=&\v(a_i)-\left[\int\limits_{a_i}^{a_f} \frac{\tilde a/a_c}{Q(\tilde a)}d\tilde a\right]\times\nonumber\\
&&
\left(
-\frac{3}{2}\Omega_M a_c \partial_{\x}\partial_{\x}^{-2} \delta(\x,a_c)-T^2[D_1](a_c)\s_1-T^2[D_2](a_c)\s_2
\right)\nonumber\\
\ee
where we dropped the $(\mathrm{res})$ subscript from $\v_\mathrm{res}$ for brevity.
The $a$ in $T[\cdot](a)$ denotes the time at which to evaluate $T[\cdot]$. Note that if one sets $\s_1$ and $\s_2$ to zero above, one recovers the standard leapfrog operators, eq.~(\ref{KDKStd}).

Time evolution between $a_0$ and $a_{n+1}$ in COLA is achieved by applying the following operator on $(\x(a_0), \v(a_0)=T[\s](a_0))$:
\be\label{ALCtimestep}
\mathrm{L}_+(a_{n+1})\left(\prod_{i=0}^n\mathrm{K}(a_{i+1/2},a_{i+1};a_{i+1})\mathrm{D}(a_i,a_{i+1};a_{i+1/2})\mathrm{K}(a_i,a_{i+1/2};a_i)\right)\mathrm{L_-}(a_{0})\ .
\ee
Here we defined 
\be\label{Lops}
\mathrm{L}_\pm(a):\ \  \v(a) \ &\mapsto&\  \v(a)=\v(a)\pm\bigg(T[D_1](a) \s_1+T[D_2](a) \s_2\bigg)\ ,
\ee
which first transforms   the initial conditions for $\v$ to the rest frame of LPT observers ($L_-$), and then adds back the LPT velocities at the end ($L_+$).

\subsubsection{Modified discretization of $T$ in COLA}\label{app:modKDK}

However, there are other options of discretizing $T$, and in all experiments we ran with COLAcode, we found the following prescription to be performing consistently better (albeit to different degrees) in terms of recovered density cross-correlation (to the truth) and halo statistics. However, we do not guarantee that this will always be the case. So, experimentation is always advised with COLA.

To make the modified discretization of $T$ used in COLA more transparent, we first apply it to a toy model. Let us say that the equation of motion for the residual displacement is
\be\label{toyEOM}
\partial^2_t \s_{\mathrm{res}}=f(\s_{\mathrm{res}},\v_{\mathrm{res}})\ , \hbox{ with }\v_{\mathrm{res}}=\partial_t \s_{\mathrm{res}}
\ee
for some $f$. The formal solutions are:
\be\label{toySol}
\s_{\mathrm{res}}(t)&=&\s_{\mathrm{res}}(t_0)+\int^t_{t_0}\v_{\mathrm{res}}(t)dt\nonumber\\
\v_{\mathrm{res}}(t)&=&\v_{\mathrm{res}}(t_0)+\int^t_{t_0}f(t)dt\ ,
\ee
with $f(t)\equiv f(\s_{\mathrm{res}}(t),\v_{\mathrm{res}}(t))$.

Now, let us assume that $\v_{\mathrm{res}}(t)$ has a time dependence which is entirely captured by some function $u(t)$, which is universal for all CDM particles. We discuss below why this choice is relevant in the cosmological context. We can then write
\be
\frac{\v_{\mathrm{res}}(t)}{u(t)}&=&\frac{\v_{\mathrm{res}}(t_c)}{u(t_c)}\ \ \ \hbox{          and} \\
\frac{f(t)}{\partial_t u(t)}&=&\frac{\partial_t\v_{\mathrm{res}}(t)}{\partial_t u(t)}=\frac{\left.\partial_t\v_{\mathrm{res}}(t)\right|_{t=t_c}}{\left.\partial_t u(t)\right|_{t=t_c}}=\frac{f(t_c)}{\left.\partial_t u(t)\right|_{t=t_c}}
\ee
for some fixed reference time, $t_c$.
We can use the above two equations to eliminate $\v_{\mathrm{res}}(t)$ and $f(t)$ from (\ref{toySol}) in favor of $\v_{\mathrm{res}}(t_c)$ and $f(t_c)$. The result is:
\be\label{toyKDK}
\s_{\mathrm{res}}(t)&=&\s_{\mathrm{res}}(t_0)+\frac{\v_{\mathrm{res}}(t_c)}{u(t_c)}\int^t_{t_0}u(t)dt\nonumber\\
\v_{\mathrm{res}}(t)&=&\v_{\mathrm{res}}(t_0)+\frac{f(t_c)}{\left.\partial_t u(t)\right|_{t=t_c}}\int^t_{t_0}\partial_t u(t)dt\nonumber\\
&=&\v_{\mathrm{res}}(t_0)+\frac{f(t_c)}{\left.\partial_t u(t)\right|_{t=t_c}}\bigg(u(t)-u(t_0)\bigg)\ .
\ee
The above solutions can be used to define Kick and Drift operators. Thus, we achieved writing down a modified discretization for the $\partial_t$ operator based on the existence of the function $u(t)$.

Now let us go back to the original cosmological problem. We will use perturbation theory to motivate the existence of $u(t)$ and arrive at a guess for it. The residual displacement $\s_{\mathrm{res}}$ is third order by construction (when 2LPT is employed for calculating $\x_{\mathrm{LPT}}$), and therefore at early times (in matter domination) $\s_{\mathrm{res}}$  has a time dependence going as $a^{n_{\mathrm{LPT}}-1/2}$, with a corresponding time dependence for $\v_{\mathrm{res}}$ going as $a^{n_{\mathrm{LPT}}}$. At third order, we have a discrete spectrum of decaying and growing modes which correspond to the following set of powers of $a$ for $\s_{\mathrm{res}}$: -9/2, -2, -3/2, 1/2, 1, 3. Depending on how we set up our initial conditions, we can excite any linear combination of these modes, and thus a realistic value of $n_{\mathrm{LPT}}$ should be a real number between $(-4)$ and $7/2$. Thus, we have arrived at an ansatz\footnote{One can envision a slight modification to our ansatz for $u(a)$ by assuming that $u(a)=A+B a^{n_{\mathrm{LPT}}}$, with $A$ and $B$ being constant, and the second term much smaller than the first. That is equivalent to setting $u=1$ in the Drift operator in (\ref{KDKALC}), while setting $u(a)=a^{n_{\mathrm{LPT}}}$ in the Kick operator. Indeed in our numerical experiments we found that such a choice is suitable for certain set-ups with initial conditions set at $z=49$ for a box at $500\,$Mpc/h. We strongly advise that one experiments with these choices to find the optimal one.} for $u(a)$: $u(a)=a^{n_{\mathrm{LPT}}}$. We discuss how we find an optimal value for $n_{\mathrm{LPT}}$ below. Before that let us write the modified Kick and Drift operators.

We can discretize (\ref{eomALC}) following the same procedure that allowed us to go from eq.~(\ref{toyEOM}) to eq.~(\ref{toyKDK}). Thus, we find:
\be\label{KDKALC}
\nonumber\mathrm{D}(a_i,a_f;a_c):& \ \  \x(a_i) \ \mapsto\  \x(a_f)=&\x(a_i)+\frac{\v(a_c)}{u(a_c)} \int\limits_{a_i}^{a_f} \frac{u(\tilde a)}{Q(\tilde a)}d\tilde a+\\
&&+(D_1(a_f)-D_1(a_i))\s_1+(D_2(a_f)-D_2(a_i))\s_2\ ,\nonumber
\\
\mathrm{K}(a_i,a_f;a_c):& \ \ \v(a_i) \ \mapsto  
\v(a_f)=&\v(a_i)+\frac{u(a_f)-u(a_i)}{T[u](a_c)}\times\nonumber\\
&&
\left(
-\frac{3}{2}\Omega_M a_c \partial_{\x}\partial_{\x}^{-2} \delta(\x,a_c)-T^2[D_1](a_c)\s_1-T^2[D_2](a_c)\s_2
\right)\nonumber\\
\ee
where we again dropped the $(\mathrm{res})$ subscript from $\v_\mathrm{res}$ for brevity.
 And again, the full time evolution is given by eq.~(\ref{ALCtimestep}).

Analogously to deriving eq.~(\ref{toyKDK}), above we used our assumption, $$\v(\tilde a)/u(\tilde a)=\v(a_c)/u(a_c)\ ,$$
to take out $\v/u$ outside the integral in the Drift operator. Similarly, for the Kick operator we used that 
\be
T[\v]&=&-\frac{3}{2}\Omega_M a\partial_{\x}\partial_{\x}^{-2}\delta(\x,a)-T^2[D_1]\s_1-T^2[D_2]\s_2=\nonumber\\
&=&\left(
-\frac{3}{2}\Omega_M a_c \partial_{\x}\partial_{\x}^{-2} \delta(\x,a_c)-T^2[D_1](a_c)\s_1-T^2[D_2](a_c)\s_2
\right)\times\frac{T[u](a)}{T[u](a_c)}\nonumber\\
&\propto& T[u](a)\ ,\nonumber
\ee
which is trivially integrated in $a$ to give the Kick operator above.

We find that a uniform distribution in $a$ for the $a_i$'s works much better for the above integration scheme than a uniform distribution in $\log( a)$. We checked this for both sets of Kick and Drift operators above, eq.~(\ref{KDKALCstd}) and eq.~(\ref{KDKALC}), for a number of initial setups.
Thus, the optimal COLA variant for the illustrative example in the paper is given by equations (\ref{ALCtimestep},\ref{Lops},\ref{KDKALC}) with timesteps uniform in the scale factor, $a$.

To find the optimal $n_{\mathrm{LPT}}$, we perform a global fit by choosing the $n_{\mathrm{LPT}}$ which produces the best cross-correlation coefficient between predicted and true density field, as well as halo statics. One may want to allow $n_{\mathrm{LPT}}$ to vary with each successive timestep. However, we find that for the 10 timesteps we perform between $z=9$ and $z=0$, only the first one or two are  significantly affected by $n_{\mathrm{LPT}}$, since for them $\Delta a/a$ is largest (as $\Delta a$ is chosen constant as described above). Thus, we use a constant $n_{\mathrm{LPT}}$, which significantly reduces the parameter space needed to be explored.

One should note that $n_{\mathrm{LPT}}$ can depend on various parameters, including the particular code used for generating the initial conditions, the choice for the initial redshift, box size, force resolution, as well as particle masses and choice between glass or grid initial conditions. Thus, one should always fit $n_{\mathrm{LPT}}$ for the particular initial set-up at hand before generating numerous realizations. For the  set-up used as an example in this paper, we find an optimal value of $n_{\mathrm{LPT}}=-2.5$, which can be changed by $\pm 0.5$ without noticeable effect on halo statistics or final density cross-correlation.

The above COLA method can be easily implemented in any existing N-body code. One needs to keep track of  the constant $\s_1$ and $\s_2$ vectors for each particle, so that they can be used in the modified Kick and Drift operators above. Those vectors are generated as a by-product of any 2LPT initial-conditions code. The only other additional piece that is needed is calculating all quantities above which depend on $a$, such as $D_1, T[D_1], T^2[D_1],$ etc, which is a straightforward exercise.

\section{Deriving $n_{\mathrm{max}}(k)$}\label{app:nmax}

In this section we motivate the expression, eq.~(\ref{nmax}), that we use for $n_{\mathrm{max}}(k)$. In analogy with \cite{2012JCAP...04..013T} let us write the true non-linear fractional overdensity, $\delta_{\mathrm{NL}}$ as
\be
\delta_{\mathrm{NL}}(\bm{k},t)=R(k,t)\delta_{\mathrm{M}}(\bm{k},t)+\varepsilon(\k,t)
\ee
where $\delta_{\mathrm{M}}$ is some model to the true density (e.g. perturbation theory or a COLA density field). We choose the residual, $\varepsilon$, to have a vanishing 2-pt function with $\delta_{\mathrm{M}}$, i.e. $\langle \delta_{\mathrm{M}}^*\varepsilon\rangle=0$. Note, however, that nothing prevents higher order cummulants between $\delta_{\mathrm{M}}$ and $\varepsilon$ to be non-zero. 

The transfer function $R(\k)$ is easily obtained by bracketing the above expression with $\delta_{\mathrm{M}}^*$ resulting in: $R=P_{\times}/P_{\mathrm{M}}$,
where $P_{\times}\equiv \langle \delta_{\mathrm{NL}}(\k)\delta_{\mathrm{M}}^*(\k)\rangle$
and $P_{\mathrm{M}}=\langle |\delta_{\mathrm{M}}(\k)|^2\rangle$. We also define $P_{\mathrm{NL}}=\langle |\delta_{\mathrm{NL}}(\k)|^2\rangle$ and $P_\varepsilon\equiv \langle|\varepsilon(\k)|^2\rangle$. In \cite{2012JCAP...04..013T} we showed that as long as $\varepsilon$ is small, one can easily extract $R$ from simulations by taking the ratio of the cross and model power spectra. Crucially we found that calculating $R$ in that way is largely unaffected by sample variance (see \cite{2012JCAP...04..013T} for details). Thus, we assume that $R$ is known in advance.

Assuming that the residual is small, to first order in $\varepsilon$ we can write the $n$-point function of the non-linear density field schematically as
\be
\langle \delta_{\mathrm{NL}}^n\rangle=R^n\langle \delta_{\mathrm{M}}^n\rangle + nR^{n-1}\langle \delta_{\mathrm{M}}^{n-1}\varepsilon\rangle+\mathcal{O}
(\varepsilon^2)
\ee
The leading term (first term on the right hand side) tells us that the relation between the $n$-point function calculated in the model and the true result is a simple deterministic scale- and time-dependent rescaling. That term requires knowing $R$, which is calculated easily with 2-point statistics, as well as $\langle \delta_{\mathrm{M}}^n\rangle$, which is a quantity which can in principle be calculated  from the model. The terms depending on the residual, however, are not calculable inside the model (or using simple $2$-point statistics as in the case of $R$), and we consider them non-trivial. 

Thus, the maximum $n$-point function, $n_{\mathrm{max}}$, that our model allows us to calculate is given by requiring that the ratio of the first order to the zero order (in $\varepsilon$) terms above is small. So, we find
\be
n\ll R\frac{\langle \delta_{\mathrm{M}}^n\rangle}{\langle \delta_{\mathrm{M}}^{n-1}\varepsilon\rangle}\sim 
R \sqrt{\frac{P_{\mathrm{M}}}{P_\varepsilon}}
\ee
In the last step above we assumed that to an order of magnitude the $n$-point functions can be written as the square root of a product of $n$ power spectra. Using the fact that $P_\varepsilon=P_{\mathrm{NL}}-R^2 P_{\mathrm{M}}$, we end up with
\be
n\ll \frac{P_\times}{\sqrt{P_{\mathrm{NL}}P_{\mathrm{M}}}}\times \frac{1}{\sqrt{1-\frac{P_\times^2}{ P_{\mathrm{NL}}P_{\mathrm{M}} }}}=\frac{\rho_{\mathrm{M,NL}}}{\sqrt{1-\rho_{\mathrm{M,NL}}^2}}
\ee
In writing the last equality we used the definition of the cross-correlation coefficient. 

Thus, the above expression gives us an order of magnitude estimate of $n_{\mathrm{max}}(k)$. To fix the overall coefficient of $2/3$ in $n_{\mathrm{max}}$ in  eq.~(\ref{nmax}), we require that $n_{\mathrm{max}}=2$ when $P_{\varepsilon}/P_{\mathrm{NL}}=10\%$. 

\bibliography{mildly_NL_v2}

\end{document}